\documentclass[12pt,aps,pra,eqsecnum,subeqn,graphicx]{article}
\usepackage{amssymb,amsmath,color,times,epsfig}
\newcommand{\su}{\rm{su}}
\newcommand{\SU}{\rm{SU}}
\newcommand{\SUSY}{\rm{SUSY}}
\newcommand{\diag}{\rm{diag}}
\newcommand{\Remot}{\mathrm{Re\,}}
\newcommand{\Imot}{\mathrm{Im\,}}

\newcommand{\be}{\begin{equation}}
\newcommand{\ee}{\end{equation}}
\newcommand{\ba}{\begin{array}}
\newcommand{\ea}{\end{array}}
\newcommand{\beqa}{\begin{eqnarray}}
\newcommand{\eeqa}{\end{eqnarray}}
\newcommand{\beqas}{\begin{eqnarray*}}
\newcommand{\eeqas}{\end{eqnarray*}}
\newcommand{\beqal}{\begin{lefteqnarray}}
\newcommand{\eeqal}{\end{lefteqnarray}}

\begin{document}

\title{\mathversion{bold}Generalized coherent and
  squeezed states based on the $h(1) \oplus  \su(2)$ algebra}

\author{Nibaldo Alvarez M.\thanks{email address: alvarez@dms.umontreal.ca} and
 V\'eronique Hussin\thanks{email address: hussin@dms.umontreal.ca}\\
\small{D\'epartement de Math\'ematiques et de  statistique and  Centre de Recherches Math\'ematiques} \\
\small{Universit\'e  de Montr\'eal, C.P. 6128, Succ.~Centre-ville,
Montr\'eal, Qu\'ebec \ H3C 3J7, Canada}}

\maketitle \thispagestyle{empty}

%\date{\today}

%\pacs{03.65.Fd, 11.30.-j, 42.50.-p.}
\thispagestyle{empty}
\begin{abstract}
States which minimize the Schr\"odinger--Robertson uncertainty relation are
constructed as eigenstates of an operator which is a element of the $h(1)
\oplus \su(2)$ algebra. The relations with supercoherent and supersqueezed
states of the supersymmetric harmonic oscillator are given. Moreover, we are
able to compute gneneral Hamiltonians which behave like the harmonic
oscillator Hamiltonian or are related to the Jaynes--Cummings Hamiltonian.
\end{abstract}

\newpage
\baselineskip 0.73 cm \setcounter{equation}{0}\section{Introduction}

Minimum uncertainty states (MUS) are usually understood through the
minimization of the
Heisenberg uncertainty relation (HUR). These
states are well-known \cite{kn:Sh1926} since long and associated with the
so-called
Coherent States
(CS) \cite{kn:KlSk} and Squeezed States (SS) \cite{kn:Taka}. But, it has
been
observed
\cite{kn:Dodo,kn:Trifo,kn:Puri} that
a more accurate uncertainty relation may be used to construct generalized CS
and SS.
Indeed, this relation known as the Schr\"{o}dinger--Robertson uncertainty
relation (SRUR) \cite{kn:Mer} can be minimized and gives rise to new classes
of CS and SS
which have received different names in the litterature, such as
correlated states \cite{kn:Dodo} or intelligent states \cite{kn:Trifo}.
There
are two main reasons to consider such last states. First, when the two
hermitian
operators entering in the SRUR are non canonical operators, i.e.\ their
commutator is not
a multiple of the identity, the HUR could be redundant
while the SRUR not. Second, the MUS that minimize the SRUR are shown to be
eigenstates of
a linear combination of the two hermitian operators entering in the SRUR.

Recently \cite{kn:Brif} a connection has been made with the CS and SS based
on
group
theoretic
approches \cite{kn:Pere} and the concept of Algebra
Eigenstates (AES). In particular, AES have been constructed for the algebras
$\su(2)$ and
$\su(1,1)$. This
concept constitute a unification of different definitions of CS and SS.

In this paper, we give a general construction of AES based on the direct sum
$h(1) \oplus
\su(2)$. The Heisenberg algebra $h(1)$ being
relevant for the problem of the harmonic oscillator and the algebra $\su(2)$
for
particles with spin, we have a procedure to find general CS and SS for
supersymmetric systems, for example. These are clearly MUS for which the
dispersions of
corresponding operators may be calculated easily. We show finally how to use
these states
in the construction of particularly relevant Hamiltonians and in the
calculation of their
dispersions.

In the Section~\ref{sec-2os}, we put the emphasis on the SRUR and its
relevancy with
respect to the determination of MUS. The application to the position and
momentum  operators MUS
leads to the well-known CS and SS of the harmonic oscillator while when the
angular
momentum operators MUS are considered we have in mind the $\su(2)$ CS and
SS.
These
particular applications are given to bring a new light on these states and
also to
facilitate the treatement of the $h(1) \oplus \su(2)$ CS and SS.
In Section~\ref{sec-3res}, we construct the AES  based on the $h(1)
\oplus
\su(2)$
algebra and show how this gives CS and SS which generalize
the supercoherent and supersqueezed states
obtained in  other approaches \cite{kn:ArZy,kn:OrSa}. Finally, in
Section~\ref{sec-4uatro}, we construct general
 Hamiltonians similar to the one of the harmonic oscillator but where the
so-called
annihilation operator is now an element
of the algebra $h(1) \oplus \su(2)$. This permits us to use our CS and SS to
compute the
mean value and the dispersions of the corresponding energies. We show also
 how the well-known Jaynes--Cummings Hamiltonian enters in this scheme.

\setcounter{equation}{0}\section{Coherent and squeezed states as
minimum uncertainty \\
 states}\label{sec-2os}

This section will be concerned by the general definition and  properties of
MUS
($\S$\ref{sec-uno}). They are explicitly constructed when the usual position
and
momentum operators are considered ($\S$\ref{sec-xpst}) as well as when the
angular
momentum operators are taken ($\S$\ref{sec-angular}).  The connection is made
with already known results.

\subsection{Minimum uncertainty relation}\label{sec-uno}

It is well-known \cite{kn:Mer} that, for two hermitian operators $ A$ and $
B$
such that
the commutator is
\begin{equation}
[A,  B ] = i  C, \quad  C \ne 0, \label{com-rel}
\end{equation}
the HUR
\begin{equation}
{(\Delta  A)}^{2}{(\Delta  B)}^2  \geq    \frac{\langle C \rangle^2}{4}
\label{relheis}
\end{equation}
is satisfied. The mean value and dispersion of a given operator $X$ are
defined, as
usual, by
\begin{equation}
\langle  X \rangle = \bigl\langle \psi |  X | \psi \bigr\rangle, \quad (
\Delta  X)^2 =
\langle  X^2 \rangle - \langle  X \rangle^2,
\end{equation}
for a normalized state $| \psi \rangle$ describing the evolution of a
quantum
system.  As
observed by Puri~\cite{kn:Puri}, for noncanonical operators,
i.e.\ such that $C$ is not a multiple of the identity $ I$, we can have
$\langle  C \rangle=0$ and the relation \eqref{relheis}  is then redundant.
The  SRUR \cite{kn:Sh1926,kn:Mer} is never redundant and writes:
\begin{equation}
{(\Delta  A)}^{2}{(\Delta  B)}^2  \geq  \frac{1}{4} \bigl( {\langle  C
\rangle}^2 +
{\langle  F \rangle}^2 \bigr), \label{relschro}
\end{equation}
where  $\langle  F \rangle$ is a measure of the correlation between $A$ and
$B$. The
operator $ F$ is hermitian and given by
\begin{equation}
 F = \bigl\{  A - \langle   A \rangle  I,  B - \langle   B \rangle  I
\bigr\},
\end{equation}
where $\{\ ,\ \}$ denotes the anti-commutator. If there is no correlation
between the
operators $ A$ and $ B$, i.e.\ if $\langle  F \rangle=0$, the
SRUR reduces to the usual HUR.

We are interested here in the description of states which minimize the SRUR
\eqref{relschro}. A necessary and sufficient condition to
get them is to solve the eigenvalues equation:
\begin{equation}
[ A + i \lambda  B ] | \psi \rangle =  \beta | \psi \rangle, \label{egalite}
\end{equation}
where
\begin{equation}
\beta = \bigl[ \langle   A \rangle + i \lambda \langle  B \rangle \bigr],
\quad \lambda
\in {\mathbb C}, \ \lambda\ne 0. \label{egalitee}
\end{equation}
Note that, if $\Remot   \lambda \ne 0$, once we know the value of $\beta$,
this last
relation may be inverted to give the mean values
\begin{equation}
\langle  A \rangle =  \Remot\beta + \frac{\Imot \lambda}{\Remot \lambda }
\Imot  \beta ,
\quad  \langle  B \rangle = \frac{\Imot \beta}{\Remot  \lambda }
\label{valmoy3}
\end{equation}
and, if  $\Remot\lambda =0$, we get
\begin{equation}
\langle  A \rangle = \Remot \beta +  \Imot \lambda  \langle   B \rangle.
\end{equation}

As a consequence of \eqref{egalite}, one has
\begin{equation}
{(\Delta  A)}^2 = |\lambda| \Delta, \quad {(\Delta  B)}^2 =
\frac{1}{|\lambda|
}\Delta,
\end{equation}
with
\begin{equation}
\Delta = \frac{1}{2} \sqrt{{\langle  C \rangle}^2 + {\langle  F \rangle}^2}
\label{productovarto}.
\end{equation}
So the states $| \psi \rangle$ satisfying \eqref{egalite} with $|\lambda|=1$
will be
called \textbf{coherent} because they satisfy
\begin{equation}
{(\Delta  A)}^2 = {(\Delta  B)}^2 = \Delta,
\end{equation}
i.e.\ the dispersions in $ A$ and $ B$ are the same and minimized in the
sense
of SRUR.
The states $| \psi \rangle$ satisfying \eqref{egalite} with
 $|\lambda| \ne 1$ will be called  \textbf{squeezed} because if $|\lambda|
<
1$, we have
$ {(\Delta  A)}^2 < \Delta < {(\Delta  B)}^2$ and
if $|\lambda| > 1$, we have $ {(\Delta  B)}^2 < \Delta < {(\Delta  A)}^2$.

Some other relations are also useful for our considerations.  The direct
computation of
${(\Delta  A)}^2$ and ${(\Delta  B)}^2$ is usually complicated but in the
MUS
that
satisfy \eqref{egalite}, we can write
\begin{eqnarray}
{(\Delta  A)}^2
        &=& \frac{1}{2} \bigl| \Remot\lambda \langle  C \rangle +   \Imot
\lambda
\langle  F \rangle \bigr| , \label{disA2}\\
{(\Delta  B)}^2
        &=& \frac{1}{2   {|\lambda|}^2 } \bigl| \Remot \lambda \langle  C
\rangle + \Imot
\lambda \langle  F \rangle \bigr| , \label{disB2}
\end{eqnarray}
with
\begin{equation}
\Imot \lambda    \langle  C \rangle =    \Remot \lambda   \langle  F
\rangle.
\label{cfrelation}
\end{equation}
For $\Remot\lambda=0$, we have $\langle  C \rangle = 0$, which corresponds
to
the case
where the HUR is redundant. The MUS satisfy the
minimum SRUR (MSRUR)
\begin{equation}
{(\Delta A)}^2 {(\Delta B)}^2 = \Delta^2,
\end{equation}
with
\begin{equation}
{(\Delta  A)}^2 = \frac{1}{2} \bigl| \Imot \lambda   \langle  F \rangle
\bigr|
,  \quad
{(\Delta  B)}^2 = \frac{1}{2}   \biggl| \frac{\langle  F \rangle}{ \Imot
\lambda} \biggr|
\end{equation}
and
\begin{equation}
\Delta = \frac{1}{2}  \bigl| \langle  F \rangle \bigr|.
\end{equation}
For $\Remot \lambda \ne 0$, from \eqref{cfrelation}, we have
\begin{equation}
\langle  F \rangle = \frac{\Imot \lambda}{\Remot  \lambda} \langle  C
\rangle.
\label{moyf}
\end{equation}
Moreover, from \eqref{disA2} and \eqref{disB2}, we get
\begin{equation}
{(\Delta  A)}^2 = \biggl| \frac{ {| \lambda|}^2}{2\Remot \lambda} \langle
C
\rangle
\biggr|, \quad
{(\Delta  B)}^2 = \biggl| \frac{1}{2\Remot \lambda} \langle  C \rangle
\biggr|
\label{moyab}
\end{equation}
and, then,
\begin{equation}
\Delta = \biggl| \frac{| \lambda|}{2\Remot \lambda} \langle  C \rangle
\biggr|.
\label{delfactor}
\end{equation}
In this case, it is sufficient to compute the mean value of $ C$ to deduce
that of $F$
and the dispersions. The particular case
where $\Imot\lambda =0$ corresponds to the fact that the MSUR co\"{\i}ncides
with the
minimum HUR (MHUR).

\subsection{Position and momentum coherent and squeezed states}
\label{sec-xpst}

Let us apply the preceding considerations to the special case of the usual
position $ x$
and momentum $ p$ operators of a given quantum system.
The canonical commutation relation (if $\hbar =1$) being
\begin{equation}
[ x, p] = i  I,
\end{equation}
the SRUR writes:
\begin{equation}
{(\Delta  x)}^{2}{(\Delta  p)}^2  \geq  \frac{1}{4} \bigl( 1 + {\langle  F
\rangle}^2
\bigr). \label{inchar}
\end{equation}
The MUS $| \psi, \lambda, \beta \rangle$ satisfy the eigenvalues equation:
\begin{equation}
[ x + i \lambda  p ] | \psi, \lambda, \beta \rangle =  \beta | \psi,
\lambda,
\beta
\rangle.  \label{ecuvapro1}
\end{equation}
If we introduce the usual creation $ a^\dagger$ and annihilation $a$
operators
\begin{equation}
 a^\dagger = \frac{ x - i  p}{\sqrt2}, \quad  a = \frac{ x + i  p}{\sqrt2},
\label{creani}
\end{equation}
such that $[ a, a^\dagger] =  I$, the equation \eqref{ecuvapro1} becomes
\begin{equation}
\frac{1}{\sqrt2} \bigl[ (1 - \lambda)  a^\dagger + (1 +\lambda)  a \bigr] |
\psi,
\lambda, \beta \rangle = \beta | \psi, \lambda, \beta \rangle.
\label{evalpaa}
\end{equation}

The general resolution of Eq.~\eqref{evalpaa} is obtained by expressing the
state $|
\psi, \lambda, \beta \rangle$ as a superposition of the energy
 eigenstates $\bigl\{ | n \rangle, n=0,1,2,\dots \bigr\}$ of the usual
harmonic
oscillator Hamiltonian
\begin{equation}
 H_0 = w \biggl( {a}^\dagger  a + \frac{1}{2} \biggr).
\end{equation}

Let us recall that these eigenstates satisfy
\begin{equation}
 a   | n \rangle = \sqrt{n}   | n-1 \rangle, \quad { a^\dagger}    | n
\rangle
= \sqrt{n
+1}  | n+1 \rangle   \label{creanni}
 \end{equation}
and we can write them as
\begin{equation}
| n \rangle = \frac{{{a}^{\dagger}}^n}{\sqrt{n!}} | 0 \rangle, \quad
n=0,1,2,\dots.
\end{equation}
So if we insert
\begin{equation}
| \psi, \lambda, \beta \rangle = \sum_{n=0}^\infty C_{\lambda,\beta, n} | n
\rangle ,
\quad  C_{\lambda,\beta, n}  \in {\mathbb C}, \label{fondaserie}
\end{equation}
in Eq.~\eqref{evalpaa}, using the expressions \eqref{creanni}, we get the
recurrence
system
$$
\frac{1}{\sqrt2} \bigl[ \sqrt{n} (1 - \lambda) C_{\lambda,\beta,n-1} +
\sqrt{n+1} (1 +
\lambda) C_{\lambda,\beta, n + 1}\bigr] = \beta  C_{\lambda,\beta,n}, \quad
n
= 1, 2,3,
\dots,
$$
\begin{equation}
\frac{(1 + \lambda)}{\sqrt2} C_{\lambda,\beta, 1}  =  \beta
C_{\lambda,\beta,0}.
\label{recucoeffgen}
\end{equation}
The case $\lambda = -1$ does not give any solution and must be eliminated.
If
we set
\begin{equation}
\biggl(\frac{1- \lambda}{1 + \lambda }\biggr) = \delta e^{i \phi}  , \quad
\delta \in
{\mathrm R}_+, \phi \in  \biggl[-\frac{\pi}{2},
\frac{3 \pi}{2}\biggr[, \label{laphi}
\end{equation}
the resolution of the  recurrence system \eqref{recucoeffgen}  leads to the
general
solution of Eq.~\eqref{evalpaa}:
\begin{equation}
| \psi, \lambda, \beta \rangle = C_{\lambda, \beta, 0}
\exp\biggl(- \delta e^{i \phi} \frac{{ a^\dagger}^2}{2} \biggr)
\exp\biggl( \frac{\beta}{\sqrt2} ( 1 + \delta e^{i \phi})  a^\dagger \biggr)
| 0 \rangle. \label{solcompose}
\end{equation}
The special case $\lambda = 1$ corresponds to $\delta = 0$ and gives rise to
the usual
expression of the CS of the harmonic oscillator.
These states \eqref{solcompose}  can also be obtained as the action of two
unitary
operators on the fundamental state. The first one  \cite{kn:Pere} is
the usual displacement operator $D$ associated with an irreducible
representation of the
Heisenberg--Weyl group $H(1)$ with algebra
$h(1) = \{ a, a^\dagger, I\}$.  The second one is the squeezed  operator
 $S$ associated  with an irreducible representation of $\SU(1,1)$ with
algebra
$\su(1,1) = \bigl\{ a^2, {(a^\dagger )}^2, a a^\dagger + a^\dagger a
\bigr\}$.
This is a
known fact  \cite{kn:workshop} when squeezed states of the harmonic
 oscillator are studied. We have explicitly
\begin{equation}
| \psi, \lambda, \beta \rangle =  S \bigl(\chi(\delta, \phi)\bigr)  D(\eta)
|0\rangle,
\end{equation}
where
\begin{equation}
 D (\eta) = \exp\ ( \eta  a^\dagger - {\bar \eta}  a ) \quad {\rm and} \quad
 S (\chi) = \exp \biggl( \chi \frac{{ a^\dagger}^2}{2} - {\bar \chi} \frac{
a^2}{2}\biggr)
\label{cosq-uni}
\end{equation}
with
\begin{equation}
\eta = \frac{\beta}{\sqrt2 } \frac{( 1 + \delta e^{i \phi})}{
\sqrt{1- \delta^2} } \quad {\rm and} \quad \chi (\delta, \phi ) = -
\tanh^{-1} (\delta)   e^{i \phi}. \label{fac-cosq}
\end{equation}
The condition for having normalizable states is that $0 \le \delta <1$. Let
us
insist
here on the fact that these SS already obtained in the literature as
eigenstates of a linear combination of $a$ and $a^\dagger $ are also MUS
such
that $ {(
\Delta x )}^2 {( \Delta p )}^2 = \Delta^2 = \bigl( 1 + \langle F
\rangle^2
\bigr)/4$.
From Eq.~\eqref{moyf} and the fact
that $\langle C \rangle =1$, we get
\begin{equation}
\langle  F \rangle = \frac{\Imot\lambda}{\Remot  \lambda} = \frac{- 2 \delta
\sin
\phi}{(1 - \delta^2)}
\end{equation}
and the factor $\Delta$ is
\begin{equation}
\Delta (\delta , \phi) = \sqrt{ \frac{1}{4} \bigl( 1 + \langle F \rangle^2
\bigr)} =
\sqrt{\frac{1}{4} + \frac{\delta^2 \sin^2 \phi}{{(1 -
\delta^2)}^2}}.\label{grandelta}
\end{equation}
Moreover, from \eqref{disA2} and \eqref{disB2}, the dispersions are
\begin{equation}
{(\Delta  x)}^2   = \frac{{|\lambda|}^2}{2|\Remot \lambda|} = \frac{(1 - 2
\delta
\cos\phi + \delta^2)}{2(1-\delta^2)} \label{xosvar}
\end{equation}
and
\begin{equation}
(\Delta  p)^2 = \frac{1}{2 |\Remot \lambda |} = \frac{(1 + 2 \delta \cos\phi
+
\delta^2
)}{2(1-\delta^2)}. \label{posvar}
\end{equation}

\begin{figure}[h]
\begin{center}\begin{picture}(31.5,21) \put(0,0){\framebox(31.5,21){}}
\put(0,2.1){\includegraphics[width=70mm]{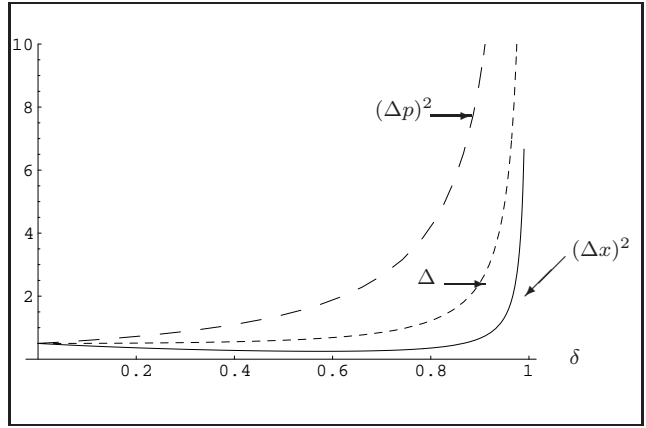}}
\put(27.9,3){\scriptsize $\delta$} \put(28,8.4){\scriptsize $(\Delta
x)^2$} \put(27.7,8.4){\vector(-1,-1){2}} \put(18.2,15.4){\scriptsize
$(\Delta  p)^2$} \put(21,15.4){\vector(1,0){2}}
\put(20.3,7){\scriptsize $\Delta$} \put(21.7,7){\vector(1,0){2}}
\end{picture}\end{center}
\caption{Graphs of the dispersions ${(\Delta  x)}^2$, ${(\Delta
p)}^2$ and the $\Delta$ factor as functions of $\delta$ for $\phi=
\pi / 6 $.} \label{fig:vari3}
\end{figure}

Let us recall now that the CS are not only the one for $\lambda=1$  but also
all the
states where $|\lambda|=1$. From the relation \eqref{laphi}, we deduce that
\begin{equation}
\lambda = \frac{1 - \delta e^{i \phi}}{1 + \delta e^{i \phi}} =
\frac{(1 - \delta^2) - 2i \delta \sin\phi}{(1 + 2 \delta \cos\phi +
\delta^2)}
\label{anda}
\end{equation}
and then
\begin{equation}
| \lambda |^2 = \frac{1 - 2 \delta \cos\phi + \delta^2}{1 + 2 \delta
\cos\phi
+ \delta^2
}. \label{modanda}
\end{equation}
This means that CS occur also for $\phi=- {\pi/2}$  or $\phi={\pi/2}$ and
$\delta \ne 0$.
The other  values of $\lambda$  describe $x$-squeezed states when
$\phi \in ]- {\pi/2}, {\pi/2}[$ and $p$-squeezed states when $\phi \in
]{\pi/2},  {3
\pi/2}[$.  On the other hand,
for fixed values of  $\phi$ the expression \eqref{grandelta} attains its
minimum value
${1/2}$ when $\delta=0$  and  when $\phi=0$ and $\phi=\pi$ for fixed values
of
$\delta$.
In the first
of these  cases, we have $\lambda=1$ and we are in  the standard CS of   the
harmonic
oscillator, i.e.\  eigenstates of the $a$ operator.  In the  second case,
$\lambda$ is a positive real quantity equal to $(1  - \delta)/(1 + \delta)
\le
1$
if  $\phi=0$ and to $(1  + \delta)/(1 - \delta) \ge 1$  if $\phi=\pi$. We
are
in
the special SS states that are eigenstates of the  $(a + \delta a^\dagger )$
and $(a -
\delta a^\dagger )$ operators respectively.

\begin{figure}[h]
\centering
\begin{picture}(31.5,21)
\put(0,0){\framebox(31.5,21){}}
\put(1,2.1){\includegraphics[width=70mm]{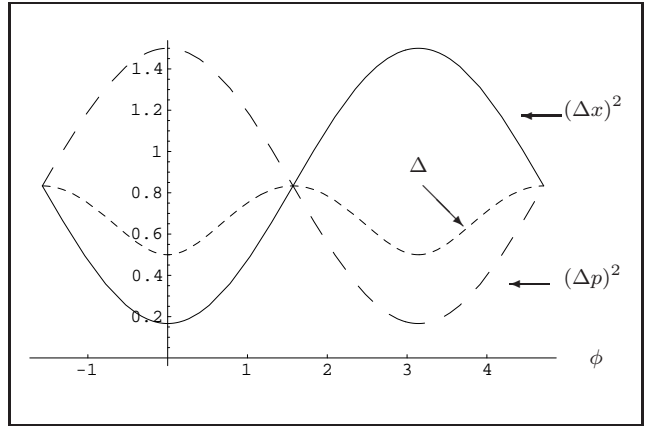}}
\put(28.9,3){\scriptsize{$\phi$}} \put(27.6,7){\scriptsize{${(\Delta
p)}^2$}} \put(26.9,7){\vector(-1,0){2}}
\put(27.6,15.4){\scriptsize{${(\Delta  x)}^2$}}
\put(27.5,15.4){\vector(-1,0){2}}
\put(19.9,12.6){\scriptsize{$\Delta$}}
\put(20.6,11.9){\vector(1,-1){2}}
\end{picture}
\caption{Graphs of the dispersions ${(\Delta  x)}^2$, ${(\Delta
p)}^2$ and the $\Delta$ factor as functions of $\phi$ for
$\delta=0.5$.} \label{fig:vari1}
\end{figure}

Fig.~\ref{fig:vari3} shows the behavior of  $(\Delta  x)^{2}$,
$(\Delta p)^2$ and $\Delta$ as functions of $\delta$ for
$\phi={\pi/6}$. In this region ${(\Delta
 x)}^{2}
\bigl((\Delta  p)^2 \bigr)$ is always less (greater) than $\Delta$,
as expected. For $\delta =0$, the three curves  co\"{\i}ncide, the
intersection point corresponds to the CS $| \psi, 1, \beta \rangle$.
The value of $\Delta =$ \eqref{grandelta} when $\delta = 0$ is also
the minimum value $1/2$ which corresponds to the MHUR.
Fig.~\ref{fig:vari1} shows the behavior of the same quantities as
functions of $\phi$ for $\delta=0.5$. The points where the three
curves intersect are the CS.

\subsection{Angular momentum coherent and squeezed
states}\label{sec-angular}

Let us now take the angular momentum operators $J_k$ for  $k=1,2,3$, which
satisfy the
usual $\su(2)$ commutations relations
\begin{equation}
[ J_k, J_l] = i \varepsilon_{klm}  J_m, \quad k,l,m=1,2,3.
\label{su2-comm-rel}
\end{equation}
Here we want to solve the eigenvalues equation
\begin{equation}
( J_1 + i \lambda  J_2 ) | \psi, \lambda, \beta \rangle= \beta | \psi,
\lambda, \beta
\rangle \label{angulaire1},
\end{equation}
where $\beta = \bigl[ \langle  J_1 \rangle + i  \lambda \langle  J_2 \rangle
\bigr]$.
On
the
contrary of the preceding example where the HUR is never redundant
(because $x$ and $p$ are canonical), here the commutator of $J_1$ and $J_2$
is
not a
multiple of the identity and then $\langle  J_3 \rangle$
may be equal to zero for some special cases. Some of these cases have been
discussed
elsewhere \cite{kn:Puri,kn:AgPu,kn:Rash,kn:WoEb}.
Here we give the general solution of the equation \eqref{angulaire1}, for
all
possible
values of $\lambda$ and $\beta$.

It would be better to work with the operators $J_{\pm} =  J_1 \pm i  J_2$
instead of
$J_1$ and $J_2$. So that the equation \eqref{angulaire1} becomes
\begin{equation}
\frac{1}{2}  \bigl[(1 + \lambda)  J_+ + (1 - \lambda)  J_- \bigr]  | \psi,
\lambda, \beta
\rangle =  \beta   | \psi, \lambda, \beta \rangle. \label{angulaire2}
\end{equation}
Using the usual complete set of angular momentum states $\bigl\{|j, r
\rangle
\bigr\}$,
$j$ integer or half-odd integer and $ r \in \bigl\{-j, - (j-1), \dots,
j-1,j\bigr\}$, we
know that
\begin{equation}
J^2 | j, r \rangle = ( J_1^2 +  J_2^2 +  J_3^2 ) | j, r \rangle =   j (j+1)
|
j, r
\rangle,
\end{equation}
\begin{equation}
J_3 | j, r \rangle = r | j, r \rangle
\end{equation}
and
\begin{equation}
J_\pm   | j , r \rangle = \sqrt{ (j \mp r) (j \pm r + 1)}  | j, r \pm 1
\rangle.
\end{equation}
This means that for each $j$ fixed, the eigenstates $| \psi, \lambda, \beta
\rangle^j$ of
Eq.~\eqref{angulaire2} may be written as
\begin{equation}
| \psi, \lambda, \beta \rangle^j = \sum_{r=-j}^{j} C^j_{\lambda,\beta,r} |
j,r
\rangle,
\quad C^j_{\lambda,\beta,r}   \in {\mathbb C}, \label{fsu2}
\end{equation}
where the coefficients $C^j_{\lambda,\beta,r}$ satisfy a recurrence system
of
the form
\begin{equation}
(1+ \lambda) \sqrt{ ( j + r) (j - r + 1)} C^j_{\lambda,\beta,r-1} + (1 -
\lambda) \sqrt{
( j - r) (j + r + 1)} C^j_{\lambda,\beta, r+1} = 2 \beta
C^j_{\lambda,\beta,r},
\label{sysangulaire}
\end{equation}
for $ r= -j, \dots,j$ and $C^j_{\lambda,\beta,j +1}=
C^j_{\lambda,\beta,-(j+1)}= 0$.

For $\lambda = \pm 1$, the unique eigenstates are $ |\psi, \pm 1, 0
\rangle^j
= | j, \pm
j \rangle$. For $\lambda \ne \pm 1$ and $\beta =0$, the recurrence relation
\eqref{sysangulaire} is solved to give
\begin{equation}
{| \psi, \lambda, 0 \rangle}^j = C^j_{\lambda,0,j}   e^{i( j \phi /
2)} \sum_{k=0}^j {(-1)}^k   \frac{{j\choose k}}{\sqrt{ {2j\choose
2k}}}  \delta^k   e^{- i(j-2k) \phi / 2}  | j , j -2k \rangle, \quad
j \quad {\rm integer} \label{quasipuri},
\end{equation}
where we have used the formula \eqref{laphi} to express $\lambda$ in terms
of
the
$\delta$ and $\phi$. It is again possible to express such a state from the
action of unitary operators associated with an irreducible representation of
a
 group
which is here $\SU(2)$. Indeed, we have
\begin{equation}
| \psi, \lambda, 0 \rangle^j =  C^j_{\lambda,0}  \exp \biggl[ - \frac{1}{2}
\ln(\delta)
J_3 \biggr] U  | j , 0 \rangle,
\end{equation}
where
\begin{equation}
U = \exp\biggl( - \frac{\pi}{4} ( e^{- i \phi / 2}  J_+ - e^{i \phi / 2}
J_-)
\biggr) .
\label{opuri}
 \end{equation}

For the  general case  $\lambda \ne \pm 1$, the analysis of the system
\eqref{sysangulaire}  shows  that for each $j$, there exists $(2j +1)$
possible values
for the eigenvalue
$\beta$, which are
\begin{equation}
\beta^j_m = m \sqrt{1- \lambda ^2}, \quad m=-j, \dots,j.  \label{betajm}
\end{equation}
If we use the relation
\begin{equation}
[ J_1 + i \lambda  J_2 ] \biggl[\exp\biggl( - \frac{1}{2} \ln(\delta)   J_3
\biggr) \,  U
 \biggr] =\biggl[\exp\biggl( - \frac{1}{2} \ln(\delta)  J_3 \biggr)   U
\biggr]
[\sqrt{1-\lambda^2}   J_3],
\end{equation}
we see immediately that the corresponding eigenstate $|\psi, \lambda,
\beta^j_m
\rangle^j$ is
\begin{equation}
|\psi, \lambda, \beta^j_m \rangle^j \equiv {|\psi, \lambda, m \rangle}^j=
C^j_{\lambda,m}
\exp\biggl[ - \frac{1}{2} \ln(\delta)   J_3 \biggr]   U   | j , m \rangle,
\quad m=-j, \dots, j,  \label{purigen}
\end{equation}
where $ U \equiv$ \eqref{opuri}. They can be written in terms of the
Jacobi polynomials as
\beqa | \psi, \lambda, m \rangle^j
         &=&  C^j_{\lambda , m}  \nonumber  \\ & \times &   \exp\biggl( - \frac{1}{2} \ln(\delta)  J_3
\biggr) e^{i m \phi / 2}   e^{- i (\phi/2)   J_3} \nonumber \\
&\times &   \sum_{r=-j}^{j} 2^r  \sqrt{ \frac{(j+r)! (j-r)!}{(j-m)!
(j+m)!}} P^{-r+m, -r-m}_{j+r} ( 0 )  | j , r \rangle.
\label{soljac1} \eeqa

In these last states, we want to compute now the mean values and
dispersions of  some operators in order to exhibit their behavior in
the CS and SS.

If $\Remot  \lambda \ne 0$, the mean values of $ J_1$ and $ J_2$ in the
states
\eqref{soljac1} are obtained using  \eqref{valmoy3}  and  \eqref{betajm}.
In terms of $\delta$ and $\phi$ as defined by \eqref{laphi}, we get
\begin{equation}
\langle  J_1 \rangle^j_m =  2 m \frac{\delta^{1/2}}{(\delta +1)} \cos
\biggl(\frac{\phi}{2}\biggr), \quad  \langle  J_2 \rangle^j_m =  2 m
\frac{\delta^{1/2}}{(\delta +1)} \sin\biggl(\frac{ \phi}{2}\biggr).
\label{moyj1j2}
\end{equation}
The relations  \eqref{moyf}--\eqref{delfactor}  applied to our case tell us
that $(\Delta J_1)^2$, $(\Delta  J_2)^2$, $\Delta$ and $\langle  F \rangle$
are all obtained from the mean value of $ J_3$, i.e.
$$
\bigl( (\Delta  J_1)^2\bigr)^j_m = \frac{|\lambda|^2}{2\Remot \lambda}
\langle
J_3 \rangle^j_m,    \quad
\bigl( (\Delta  J_2)^2 \bigr)^j_m = \frac{1}{2 \Remot \lambda} \langle  J_3
\rangle^j_m
$$
\begin{equation}
\Delta^j_m = \frac{|\lambda|}{2 \Remot \lambda} \langle  J_3 \rangle^j_m,
\quad \langle   F  \rangle^j_m  = \frac{\Imot \lambda}{\Remot \lambda}
\langle  J_3 \rangle^j_m. \label{dispersions}
\end{equation}
The mean values of $J_3$ in the states \eqref{soljac1} or equivalently in
the
states
\eqref{purigen}  are given by
\begin{equation}
\langle  J_3 \rangle^j_m = - \frac{\partial}{\partial q} \ln
\bigl(\langle j ,m| U^\dagger e^{-q  J_3}  U |j,m\rangle \bigr),
\end{equation}
where $q= \ln \delta$. After some computations, we get
\begin{equation}
\langle  J_3 \rangle^j_m = - |m|   \tanh \biggl(\frac{ q}{2}\biggr) -
\frac{1}{2}
\sinh(q)   \bigl(j + |m| + 1\bigr)   \frac{P^{1, 1 + 2 |m| }_{j - |m| -1}
(\cosh
q)}{P^{0,  2|m|}_{j - |m|} ( \cosh q)} \label{moyj3f}.
\end{equation}
Inserting \eqref{moyj3f} into the expression \eqref{dispersions}, we get
\begin{subequations}
\begin{align}
\bigl( (\Delta  J_1)^2\bigr)^j_m = (1 - 2 \delta \cos\phi +
\delta^2)\Lambda^j_m (\delta), \quad \bigl( (\Delta  J_2
)^2\bigr)^j_m  = (1 + 2 \delta \cos\phi + \delta^2 )\Lambda^j_m
(\delta), \label{disj12} \\
 (\Delta)^j_m  = \sqrt{1- 2 \delta^2 \cos(2 \phi) + \delta^4} \Lambda^j_m
(\delta),
\quad
\langle { F} \rangle^j_m =  - 4 \delta \sin\phi  \Lambda^j_m (\delta),
\label{moyfj1j2f}
\end{align}
\end{subequations}
where
\begin{equation}
\Lambda^j_m (\delta)= \biggl[ \frac{|m|}{2( 1 + \delta)^2} + \frac{(j + |m|
+
1)}{8
\delta}\frac{{P^{1, 1 + 2 |m| }_{j - |m| -1}  ((1 + \delta^2)/2
\delta)}}{{P^{0,
2|m|}_{j - |m|}((1 + \delta^2)/2 \delta)}} \biggr].\label{factor}
\end{equation}

 The case $\Remot  \lambda=0$ may be obtained as the limit case of the
preceding one by taking  $\delta=1$ in the
 expressions \eqref{disj12}, \eqref{moyfj1j2f} and \eqref{factor}. Let us
recall that it
corresponds to
$\langle J_3 \rangle=0$ and $\lambda=-i \tan  {\phi/2} $. We get
\begin{subequations}
\begin{equation}
\bigl(( \Delta J_1)^2\bigr)^j_m =  \frac{1}{2}\bigl[ j(j + 1)-m^2 \bigr]
\sin^2
\biggl(\frac{\phi}{2}\biggr),  \quad
\bigl( (\Delta  J_2 )^2\bigr)^j_m = \frac{1}{2} \bigl[j (j + 1) - m^2 \bigr]
\cos^2\biggl(\frac{\phi}{2}\biggr), \label{2.64a}
\end{equation}
\begin{equation}
(\Delta)^j_m  = \frac{1}{4}\bigl[j(j + 1)- m^2 \bigr] |\sin\phi|  \quad
\text{and}
\quad \langle F \rangle^j_m  = - \frac{1}{2} \bigl[ j(j + 1) - m^2 \bigr]
\sin\phi, \label{2.64b}
\end{equation}
\end{subequations}
using the fact that
\begin{equation}
P^{\alpha , \beta}_n (1)  = \frac{(\alpha +1) (\alpha +2) \dots
(\alpha + n)}{n!}.
\end{equation}
These are exactly the results given by Puri~\cite{kn:Puri}.

To illustrate  these considerations by a concrete example, let us take the
``spin--$1/2$'' case, i.e.\  $j=1/2$. The
expressions \eqref{disj12}, \eqref{moyfj1j2f} thus reduce to
\begin{equation}
\bigl( (\Delta  J_1)^2 \bigr)_{\pm} = \frac{(1 - 2 \delta \cos\phi +
\delta^2
)}{4(1+
\delta)^2}, \quad \bigl( (\Delta  J_2)^2 \bigr)_{\pm} = \frac{(1 + 2 \delta
\cos\phi +
\delta^2 )}{4( 1 + \delta)^2}
\end{equation}
and
\begin{equation}
\Delta_{\pm} (\delta, \phi)  = \frac{1}{4} \sqrt{1+ 4 \biggl( \frac{
\delta^2
\sin^2 \phi
- \delta(1+ \delta)^2}{(1+\delta)^4} \biggr)}, \label{d-fac}
\end{equation}
where we have used the $\pm$ sign for the values of $m=\pm {1/2}$. The MSRUR
thus writes
\begin{equation}
\bigl( (\Delta  J_1)^2\bigr)_\pm \bigl( (\Delta  J_2 )^2\bigr)_{\pm}
=  (\Delta_{\pm} )^2 (\delta, \phi) = \frac{1}{16} \biggl[ 1+ 4 \biggl( \frac{
\delta^2 \sin^2
\phi - \delta(1+ \delta)^2}{(1 + \delta)^4}\biggr) \biggr].
\label{j1j2-dfac}
\end{equation}

\begin{figure} \centering
\begin{picture}(31.5,21)
\put(0,0){\framebox(31.5,21){}} \put(0,2.1){
\includegraphics[width=70mm]{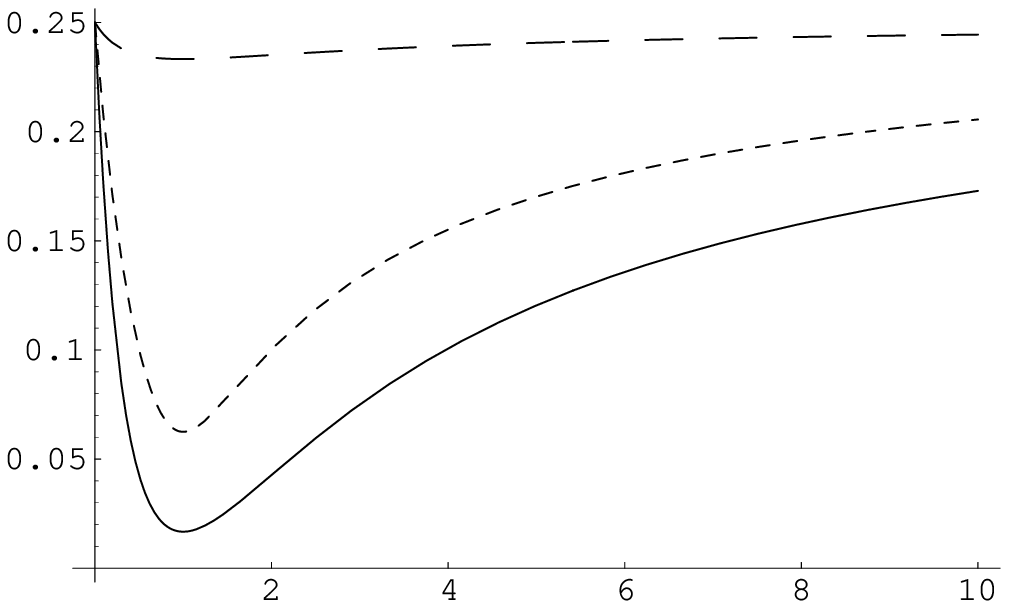}}
\put(27.9,3){\scriptsize{$\delta$}}
\put(19,8.4){\scriptsize{${\left({\left(\Delta
J_1\right)}^2\right)}_{\pm}$}}
 \put(18.4,8.4){\vector(-1,1){2}}
\put(25.6,19){\scriptsize{$ {\left({\left(\Delta
J_2\right)}^2\right)}_{\pm}$}} \put(25.4,19){\vector(-1,-1){1.5}}
\put(5.6,12.6){\scriptsize{${\Delta}_{\pm}$}}
\put(6.3,11.9){\vector(1,-1){2}}
\end{picture} \caption{Graphs of the dispersions ${\left({\left(\Delta
J_1\right)}^2\right)}_{\pm}$, ${\left({\left(\Delta
J_2\right)}^2\right)}_{\pm}$ and the ${\Delta}_{\pm}$ factor as
functions of $\delta$ for  $\phi= \pi /6 $ and $j= 1 / 2$.}
\label{fig:vari4}
\end{figure}
For fixed values of  $\phi \ne 0$ and $\pi$, the expression
\eqref{d-fac} attains its minimum value $\bigl|\sin \phi |\big/ 8$
when $\delta=1$. On
 the other hand, for fixed values of  $\delta$ such that $\delta \in  [0,1[
\cup
]1,\infty]$, the minimum of  \eqref{d-fac}  is  $(1/4) \sqrt{   \bigl[ 1 -
(4
\delta)/(1+ \delta)^2} \bigr]$ when $\phi=0$ or $\phi=\pi$. In the first
case
we have
$\lambda = - i(\sin\phi)/(1+ \cos\phi)$,
which means that we have some special classes of SS from which we recognize
CS
with
$\lambda = - i$ (eigenstates of the $J_1 + J_2$ operator) and with $\lambda
=i$
(eigenstates of the  $J_1 - J_2$ operator). In the second case, we have
$\lambda = (1  - \delta)/(1 + \delta) \le 1$ if  $\phi=0$ and $\lambda = (1
+
\delta )/(1 -
\delta) \ge 1$  if $\phi=\pi$, i.e.\  the minimum $\Delta_{\pm} (\delta,0)
=\Delta_{\pm}(\delta,\pi)$ values  occur for the special states which are
eigenstates of
the operators $(J_+ + \delta J_- )$ and $(J_+ - \delta J_- )$ respectively.
Let us recall
that the CS with $\lambda=1$ occur when $\delta =0$ and  those with
$\lambda=-1$ when
$\delta \mapsto \infty$. They correspond to the eigenstates of  $J_+$ and
$J_-$ operators
respectively.
For  such  states, according to equation \eqref{j1j2-dfac}, we have
$\bigl( (\Delta  J_1)^2\bigr)_{\pm}= \bigl( (\Delta  J_2 )^2
\bigr)_{\pm}= \bigl(\Delta_{\pm} (0, \phi)\bigr)^2= \lim_{\delta \mapsto
\infty} \bigl(\Delta_{\pm} (\delta, \phi)\bigr)^2=  {1/4}$.

\begin{figure}[h]
\centering
\begin{picture}(31.5,21)
\put(0,0){\framebox(31.5,21){}}
\put(0,2.1){\includegraphics[width=70mm]{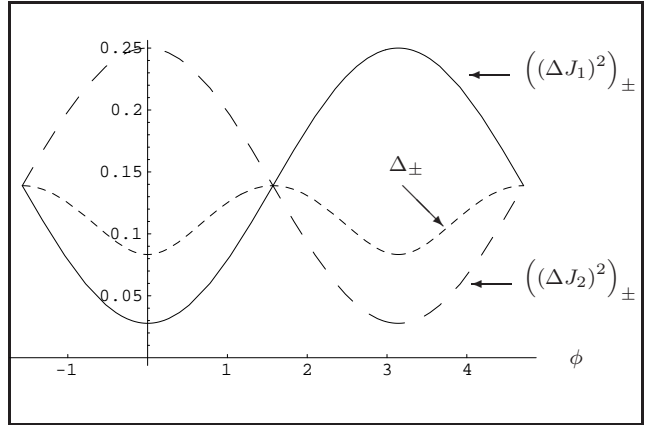}}
\put(27.9,3){\scriptsize{$\phi$}} \put(25.6,7){\scriptsize{$
{\left({\left(\Delta  J_2\right)}^2\right)}_{\pm}$}}
\put(25,7){\vector(-1,0){2}}
\put(25.6,17.4){\scriptsize{${\left({\left(\Delta
J_1\right)}^2\right)}_{\pm} $}} \put(24.9,17.4){\vector(-1,0){2}}
\put(18.9,12.6){\scriptsize{${\Delta}_{\pm}$}}
\put(19.6,11.9){\vector(1,-1){2}}
\end{picture}
\caption{Graphs of the dispersions ${ \left( {\left(\Delta
J_1\right)}^2\right)}_{\pm}$, ${\left({\left(\Delta
J_2\right)}^2\right)}_{\pm}$ and the ${\Delta}_{\pm}$ factor as
functions of $\phi$ for $\delta=0.5$ and $j= 1 / 2$.}
\label{fig:vari2}
\end{figure}

Fig.~\ref{fig:vari4} shows the behavior of the dispersions $\bigl(
(\Delta J_1 )^2\bigr)$, $\bigl((\Delta  J_2)^2\bigr)$ and $\Delta$
as functions of $\delta$ for $\phi={\pi/6}$ and $j={1/2}$. The
minimum value of $\Delta_{\pm}$ is here $0,0625$.  In
Fig.~\ref{fig:vari2}, we see that the graphs as a function of $\phi$
are very similar to ones for the preceding example of $x$ and $p$.

\setcounter{equation}{0}\section{Algebra Eigenstates Associated to
{\boldmath $\lowercase{h}(1) \oplus \su(2)$}} \label{sec-3res}

This section begins ($\S$\ref{sec-susyhar}) with a review of the SUSY
harmonic
oscillator and its Super-Coherent States (SCS) studied
by Aragone and Zypman \cite{kn:ArZy}. We follow  ($\S$\ref{sec-algdestates})
by the general construction of  AES based on the algebra $h(1) \oplus
\su(2)$.
These states are defined as eigenstates of an arbitrary linear combination of
the generators of the considered algebra \cite{kn:Brif}. Then we consider
special solutions to CS and SS for the so-called super-position and
super-momentum operators ($\S$\ref{sec-cosq-spsm}).

\subsection{The SUSY harmonic oscillator and its super-coherent states}
\label{sec-susyhar}

Let us recall that the quantum SUSY harmonic oscillator is defined as a
combination of a
bosonic and a fermionic oscillators. Its Hamiltonian is given by
\begin{equation}
{H}_{\SUSY}= w  (a^\dagger  a -  f^\dagger  f),
\end{equation}
where the bosonic creation and annihilation operators $ a^\dagger$ and $ a$
are defined
as in  \eqref{creani}  and the corresponding fermionic operators
$ f^\dagger$ and $ f$ are defined as
\begin{equation}
 f^\dagger = \sigma_+ = \frac{1}{2} (\sigma_1 + i \sigma_2), \quad  f  =
\sigma_- =
\frac{1}{2} (\sigma_1 - i \sigma_2),
\end{equation}
(the $\sigma_i$,  $i=1,2$ being the usual Pauli matrices) for the spin ${1/
2}$
fermion. We can thus write
\begin{equation}
{H}_{\SUSY}= w \biggl(a^\dagger  a - \frac{1}{2} \biggr) - \frac{w}{2}
\sigma_3.
\end{equation}

The representation space, we are working with in this context, is nothing
else than the direct product
\begin{eqnarray}
{\mathcal F} = {\mathcal F}_b \otimes {\mathcal F}_f
        &=&   \bigl\{ |n \rangle,   n=0,1,2,\dots\bigr\}
\otimes \biggl\{ \biggl| \frac{1}{2}, \frac{1}{2} \biggr\rangle=  |+
\rangle,
\biggl|\frac{1}{2}, \frac{-1}{2} \biggr\rangle=|-\rangle \biggr\}
\nonumber\\
        &=& \bigl\{ |n,+\rangle, |n,-\rangle,  n=0,1,2,\dots \bigr\}.
\label{Fock-space}
\end{eqnarray}
Following Aragone and Zypman~\cite{kn:ArZy},  SCS  may be constructed as
eigenstates of
a SUSY annihilation operators
\mbox{$\bigl( \sqrt{2}( a + \sigma_+)\bigr)$.} They are shown to be given  as a
linear combination of the
following normalized pure states
\begin{equation}
| \psi \rangle_+ = D \biggl(\frac{z}{\sqrt{2}}\biggr) |0, +\rangle
%\label{psi+}
\label{az+}
\end{equation}
and
\begin{equation}
| \psi \rangle_- =   D \biggl( \frac{z}{\sqrt{2}}\biggr) \frac{[
a^\dagger | 0,+\rangle - |0,-\rangle ]}{\sqrt{2}},
%\label{psi-}
\label{az-}
\end{equation}
in terms of the displacement operator $D$ given in  \eqref{cosq-uni}
and where we recognize in  \eqref{az+}, the usual CS of the harmonic
oscillator. A discussion
 \cite{kn:ArZy,kn:OrSa} of the properties of such states has led to the
observation that, except for the state $|\psi_+\rangle\equiv$
\eqref{az+}, no other linear combination of \eqref{az+}  and
\eqref{az-}  will minimize the usual HUR. This means that these
states satisfy $(\Delta  x)^2 (\Delta p)^2 \ge {1/4},$ the equality
between the position $x$ and the momentum $p$ being realized only
for $|\psi_+\rangle\equiv$  \eqref{az+}.

Such a fact can be clarified from our discussion of
Section~\ref{sec-uno}. The SCS \eqref{az+}  and  \eqref{az-}  are in
fact MUS for the SRUR  \eqref{relschro}  with
\begin{equation}
 A = \frac{1}{\sqrt{2}} \bigl[  ( a^\dagger +  a) + \sigma_1  \bigr] =
\biggl[  x  + \frac{\sigma_1}{\sqrt2} \biggr] \quad  \text{and}  \quad
 B =   \frac{1}{\sqrt{2}} \bigl[i  ( a^\dagger -  a)   + \sigma_2  \bigr] =
\biggl[  p + \frac{\sigma_2}{\sqrt2}\biggr], \label{AB-SUSY}
\end{equation}
these operators being different from $x$ and $p$. The SCS  are coherent in
the
sense that they satisfy the Eq.~\eqref{egalite} with $\lambda=1$.

Clearly, in such a context, through the group theory level, we are combining
the
information coming from both the Heisenberg--Weyl
$h(1)$ and the $\su(2)$ algebras realized in terms of the Pauli matrices in
the spin
${1/2}$ case. Its is then natural to ask the questions of determining
general
CS and SS for the direct sum $h(1) \oplus \su(2)$  which will indeed include
the special SCS we just discussed.

\subsection{Algebra eigenstates}\label{sec-algdestates}

We are working with the $h(1) \oplus \su(2)$ algebra generated by $ \{a,
{a}^{\dagger},  I; { J}_+ ,  J_- ,  J_3 \}$
as defined in the preceding sections.  AES~\cite{kn:Brif} for this algebra
are
defined
as eigenstates corresponding to a complex combination of the associated
generators.
A general hermitian operator $ A$ constructed from a combination of these
generators is
\begin{equation}
 A = A_1  a + \bar A_1 { a}^{\dagger} + A_2 I + A_3  J_+ + \bar A_3  J_- +
A_4
 J_3, \quad A_2,A_4 \in {\mathbb R}, \ A_1, A_3 \in {\mathbb C}.
\label{general}
\end{equation}
Two such operators, called $ A$ and $ B$, satisfy the commutation relation
 \eqref{com-rel}  with
\begin{equation}
 C =   \bigl[ i ( \bar A_1 B_1 - A_1   \overline{B}_1 )  I + 2 i ( B_3 \bar
A_3 - \overline{B}_3 A_3 )   J_3  +
i ( A_3 B_4 - A_4 B_3)  J_+  + i( A_4 \overline{B}_3 - \bar A_3 B_4)
J_- \bigr]. \label{Cgen}
\end{equation}

Once we search for states satisfying \eqref{egalite}, i.e.\  for eigenstates
of $ A + i \lambda  B$ $(\lambda \in {\mathbb C},  \lambda \ne 0)$, we are
in
fact considering AES and we know from Section~\ref{sec-uno} that they
minimize
the SRUR  \eqref{relschro}. Let us then study the
solutions of such a general eigenstate equation \eqref{egalite} for $A$ and
$B$ on the form \eqref{general}.

It is convenient to rewrite  this equation as
\begin{equation}
[\alpha _-  { a} + {\alpha_+} { a}^\dagger + \alpha_3  I + \beta_-   J_+ +
\beta_+  J_- +
\beta _3  J_3]  | \psi \rangle = z | \psi \rangle \label{promo},
\end{equation}
where
\begin{eqnarray}
\nonumber
\alpha_- &=& A_1 + i \lambda B_1, \quad \alpha_+ = \bar A_1 + i \lambda
\overline{B}_1, \quad
\alpha_3 = A_2 + i \lambda B_2,  \\
\beta_- &=& A_3 + i \lambda B_3, \quad \beta_+ = \bar A_3 + i \lambda
\overline{B}_3, \quad
\beta _3 = A_4 + i \lambda B_4.
\label{coeff}
\end{eqnarray}

 To solve \eqref{promo}, we express $ | \psi \rangle$ as a superposition of
fundamental
states $|n; j,m \rangle$ which constitute a generalization of the Fock space
 \eqref{Fock-space} for spin $j$. We write
\begin{equation}
| \psi \rangle^j = \sum_{m=-j}^{j} \sum_{n=0}^{\infty} C^j_{n,m} | n;
j,m\rangle,
\label{fondaa}
\end{equation}
for fixed $j$, integer or half-odd integer. Let us recall that we have
\begin{eqnarray}
\nonumber
 a   | n ; j , m \rangle
        &=& \sqrt{n}  | n-1 ; j , m \rangle, \\
 \nonumber
 a^\dagger  | n ; j , m \rangle
        &=& \sqrt{n+1}  | n+1 ; j , m \rangle, \\
 J_\pm  | n ; j , m \rangle
        &=&\sqrt{( j \mp m)( j \pm m +
1)}| n ; j  , m \pm 1 \rangle,
\label{actions}
\end{eqnarray}
with
\begin{equation}
\langle n ; j ,m \mid l ; j , r \rangle = \delta_{nl}  {\delta}_{mr}.
\label{orthogonalite}
\end{equation}

Inserting  \eqref{fondaa}  into  \eqref{promo}  and taking into account the
relations
 \eqref{actions}  and  \eqref{orthogonalite}, we get a recurrence
system which becomes more and more complicated as $j$ increases.  We also
notice that
the case where $\alpha_- = 0$ with $\alpha_+ \ne 0$ does not give any
solution and must
be eliminated.
Here two ways of solving it
completely are presented. The first one uses the results
obtained in Section~\ref{sec-xpst} and Appendix~\ref{sec-su2ae}  where AES
of
$\su(2)$
are explicitly constructed. It is described explicitly in this section using
operators
acting on a fundamental state. The second one is based on the method of
resolution of a
first
order system of linear differential equations and is described in the
Appendix~\ref{sec-auno}.

With respect to the discussion in Appendix~\ref{sec-su2ae}, we have mainly
two
types of
eigenvalues for $z$. The first type is given by
\begin{equation}
z= \rho^j_m + \alpha_3 + m b, \quad \rho^j_m \in {\mathbb C}, \label{rro}
\end{equation}
for fixed $j$ and where $m=-j, \dots , j$ and
\begin{equation}
b= \sqrt{4 \beta_+ \beta_- + \beta_3^2} \ne 0. \label{b-factor}
\end{equation}
If we compare the equations  \eqref{evalpaa}  and \eqref{epa2}  and their
respective
solutions \eqref{solcompose}  and  \eqref{fonor-eff}, we find the set of
solutions
\begin{equation}
|\psi \rangle^j_m =  {( C^j_m)}^{-1 /2}   \exp\biggl[- \frac{\alpha_+
}{2 \alpha_-} { a^\dagger}^2  +  \frac{ \rho^j_m}{\alpha_-}  a^\dagger
\biggr]
 T_{\mathrm{eff}}  | 0 ; j ,m \rangle \label{fon1},
\end{equation}
when $\alpha_-\ne 0$. Here   $ T_{\mathrm{eff}}$ is given by  \eqref{teff}
when
$\{\beta_+\ne 0, \beta_-\ne 0 \}$,  \eqref{teff-b+=0}
when $ \{\beta_+=0, \beta_3\ne 0 \}$,
 \eqref{teff-b-=0}   when $\{\beta_-=0,  \beta_3\ne 0 \}$ and finally the
identity when
$\{\beta_- = \beta_+ = 0, \beta_3\ne 0 \}$.

The second type corresponds to the so-called degenerate case $(b=0)$
where $ z = \rho + \alpha_3$. The sets  of independent  solutions
are now  given by \beqa  |\psi\rangle^j_{m}
          &=&  {( {C}^j_m)}^{-1 /2}  \exp\biggl[ - \frac{\alpha_+}{2
\alpha_-} { a^\dagger}^2 + \frac{ \rho}{\alpha_-}  a^\dagger \biggr] \nonumber \\
\label{solb++b3=0} & & \times\sum_{k=0}^{j-m} {(-1)}^k {j-m \choose
k}\frac{(2j-k)!}{(2j)!} {( a^\dagger)}^{j-m-k} {\biggl(
\frac{\alpha_-  J_-}{\beta_-}\biggr)}^{k} |0; j, j  \rangle \eeqa
when $ \beta_+=\beta_3=0$, \beqa |\psi\rangle^j_{m}
        &=& {( {C}^j_m)}^{-1 /2}  \exp\biggl[ - \frac{\alpha_+}{2
\alpha_-} { a^\dagger}^2 + \frac{ \rho}{\alpha_-}  a^\dagger \biggr] \nonumber\\
\label{solb--b3=0} && \times \sum_{k=0}^{j-m} {(-1)}^k {j-m \choose
k}\frac{(2j-k)!}{(2j)!} {( a^\dagger)}^{j-m-k}
{\biggl(\frac{\alpha_-  J_+}{\beta_+}\biggr)}^{k}
  |0; j ,  -j \rangle,
\eeqa when $ \beta_-=\beta_3=0$ and \beqa |\psi\rangle^j_m
        &=& {(C^j_{m})}^{-1/2} \exp\biggl[ - \frac{\alpha_+}{2
\alpha_-} { a^\dagger}^2 + \frac{ \rho}{\alpha_-}  a^\dagger \biggr] \nonumber\\
  && \times \biggl[\sum_{k=0}^{j-m} {(-1)}^k
{j-m \choose k}   \frac{(2j-k)!}{(2j)!}{(a^\dagger)}^{j-m-k}
{\biggl( \frac{\alpha_-}{ \beta_+}\biggr)}^k \frac{d^k e^{\vartheta
J_+}}{d{\vartheta }^k} \biggr] | 0; j,   -j\rangle,
\label{solb+b-b3=0} \eeqa when $\beta_+, \beta_-$ and $\beta_3$ are
different from zero and for $\vartheta = {\beta_3/(2\beta_+)} = - {2
\beta_-/\beta_3}$.

\subsection{Coherent and squeezed states for the super-position  and
\\ super-momentum  operators}\label{sec-cosq-spsm}

Let us consider the eigenstates of equation \eqref{promo} corresponding to
the
following special values of the parameters
\begin{equation}
A_4=B_4=A_2=B_2 = 0, \quad
A_1 =i B_1 = \frac{\mu}{\sqrt2}, \quad ( \mu \ne 0),  \quad A_3 = i B_3
=\frac{\tau}{\sqrt2},
\label{copa}
\end{equation}
so that $A$ will be called the super-position operator denoted by $X$ and $
B$ the
super-momentum operator denoted by $ P$. We have
\begin{equation} \label{suxp}
X
        = \frac{1}{\sqrt2} \bigl[ (\mu  a + \bar \mu  a^\dagger  ) +
 ( \tau
J_+ + \bar \tau  J_ - ) \bigr], \quad
P
        = \frac{i}{\sqrt2} \bigl[ (\bar \mu  a^\dagger -  \mu  a ) +
( \bar
\tau  J_-  -  \tau  J_+) \bigr].
\end{equation}
We see that the operators  \eqref{AB-SUSY} associated to the SCS are then a
special case
where $\mu = {\bar \mu}= \tau = {\bar \tau} = 1$ in the spin-${1/2}$
case.

The eigenstates equation \eqref{promo}  now writes
\begin{equation}
[ X + i \lambda  P ]   | \psi \rangle =  z   | \psi \rangle \label{XPL}
\end{equation}
and the operator $ C$ in \eqref{Cgen} is diagonal and takes the form
\begin{equation}
 C = | \mu |^2  I + 2  | \tau|^2    J_3. \label{cxp}
\end{equation}
Since, we have
\begin{eqnarray}
\nonumber
\alpha_-
        &=& \frac{{\mu  (1 + \lambda  )}}{\sqrt2 },
\quad
\alpha_+ = \frac{{\bar \mu  (1 - \lambda  )}}{\sqrt2 },
\quad \alpha_3 = 0,  \\
\beta_-
        &=& \frac{{\tau (1 + \lambda )}}{\sqrt2 },
\quad
\beta_+ = \frac{{\bar \tau  (1 - \lambda )}}{\sqrt2 },
\quad \beta _3 = 0
\label{coeffxpl}
\end{eqnarray}
and finally
\begin{equation}
b = \sqrt2  | \tau| \sqrt{1 - \lambda^2},
\end{equation}
we can use the preceding solutions to give  all the solutions of equation
\eqref{XPL}.

For $\lambda = 1$, we have  $\alpha_+=\beta_+ = b=0$  and the eigenstate
equation is
\begin{equation}
[ \mu  a + \tau  J_+ ]  |\psi  \rangle  = \frac{ z}{\sqrt2}  |\psi
\rangle.
\end{equation}
The normalized  solutions are obtained from  \eqref{solb++b3=0}  and take
the
form
\begin{equation}
| \psi \rangle^j_m
        =  \bigl(C^j_m (\mu, \tau) \bigr)^{-1 / 2 }  D \biggl( \frac{
z}{\mu
\sqrt{2}} \biggr)
         \biggl[  \sum_{k=0}^{j-m}
{(-1)}^k
{{j-m}\choose k } \frac{(2j-k)!}
        { (2j)!} (a^\dagger)^{j-m-k}
 \biggl(\frac{\mu  J_-}{\tau}\biggr)^k   \biggr]  | 0 ; j , j \rangle,
\label{fon2xp}
\end{equation}
where the normalization constant is given by
\begin{equation}
{C}^j_m (\mu, \tau ) = (j-m)!\biggl[  \sum_{k=0}^{j-m}  {{j-m}\choose k }
\frac{(
2j-k)!}{(2j)!} \biggl( \frac{|\mu|^2}{|\tau|^2}\biggr)^k   \biggr].
\end{equation}

Let us recall that in this case we have CS for which
\begin{equation}
(\Delta  X) = (\Delta  P) = \Delta = \frac{1}{2} \langle  C \rangle.
\label{dxdptaumu}
\end{equation}
The mean value of  $C$ is easy to compute and we have
\begin{equation}
\langle  C {\rangle}^j_m = {|\mu|}^2 + 2  {|\tau|}^2
\biggl[
j +   {|\tau|}^2  \frac{\partial}{\partial |\tau|^2}
\ln\bigl( C^j_m (\mu, \tau ) \bigr) \biggr]. \label{cmoyenne1}
\end{equation}

In the special case $j = {1/2}$, we find the normalized and orthogonal
states
\begin{equation}\label{junmedio}
| \psi \rangle^+
         =   D\biggl(  \frac{z}{\mu \sqrt{2}} \biggr)   | 0 ; + \rangle,
\quad
| \psi \rangle^-
         =   D\biggl(  \frac{z}{\mu \sqrt{2}} \biggr)
\frac{|\tau|}{\sqrt{|\mu|^2 + |\tau|^2}}
\biggl[a^\dagger | 0 ; + \rangle - \frac{\mu}{\tau}  | 0 ; - \rangle
\biggr],
\end{equation}
where $D$ is again given by  \eqref{cosq-uni}. In those states, we have
\begin{equation}
\langle  C \rangle^+ = |\mu|^2 + |\tau|^2,    \quad
 \langle  C \rangle^- = \biggl[ \bigl(|\mu|^2 +  |\tau|^2 \bigr) -
\frac{2|\mu|^2|\tau|^2}{(|\mu|^2 + |\tau|^2) } \biggr].
\label{ctaumu}
\end{equation}
This is clearly a generalisation of SCS considered by Aragone and
Zypman~\cite{kn:ArZy} and recalled in
 \eqref{az+}  and \eqref{az-}.

From \eqref{ctaumu}, we see that the dispersions of $\Delta X$ and $\Delta
P$ given by \eqref{dxdptaumu} computed in the CS $| \psi \rangle^-$ are
smaller
than
in the states $| \psi \rangle^+$.
The states $| \psi \rangle^-$ thus are the closest to classical states for
the
SUSY
harmonic oscillators (this means with respect to the super-position and
the super-momentum) while $| \psi \rangle^+$ are indeed  the ones closest to
classical
states of the standard harmonic oscillator (i.e.\ they minimize the HUR for
$ X$
and $ P$).
Let
us mention that if we take $\mu=1$, we see that $\langle C \rangle^+$ has
its
minimum
value equal to $1$ for {\bf $\tau \mapsto  0$} and in this case $X=x$ and
$P=p$. For the
same
value of $\mu$, we see that $\langle C \rangle^-$ takes the form
\begin{equation}
\langle C \rangle^- = \frac{1 + |\tau|^4}{1 + |\tau|^2},
\end{equation}
which has a minimum value $\langle C \rangle^-_{\min} = 2 (\sqrt{2} -1) <1$
for
$ {|\tau|}^2 = \sqrt{2} -1$.

For $\lambda \ne \pm 1$, from equation \eqref{fon1}  and
$T_{\mathrm{eff}}\equiv$ \eqref{opt2}), using  also \eqref{cosq-uni}
and \eqref{fac-cosq},  we get the states \beqa \nonumber | \psi
\rangle^j_m
       &=& {(C^j_m)}^{-1/2}  S \bigl( \chi (\delta,
\phi - 2
{\phi}_u )\bigr)
 D\bigl( \eta_m (z,\delta,\phi, \mu,\tau) \bigr) \\
        && \times  \exp\biggl( \frac{{-
\tau \delta^{-1/2}
e^{- i
\phi /2}}}{{ |\tau|}} J_+\biggr)
\exp\biggl( \frac{\bar \tau \delta^{1/2} e^{i \phi /2}}{2|\tau|}
J_-\biggr)
 | 0; j , m \rangle, \label{fon1xp}
\eeqa where
\begin{equation}
\eta_m ( z , \delta, \phi, \mu, \tau ) = \frac{1}{\mu}  \biggl\{ \frac{ {z (
1
+
\delta e^{i \phi})}}{\sqrt{2}}
- 2 m |\tau| \delta^{1/2} e^{i \phi /2} \biggr\}, \quad \mu = |\mu| e^{i
\phi_u}
\end{equation}
and where we have used instead of $\lambda$ the parameters $\delta$ and
$\phi$
as given
in  \eqref{laphi}.  Let us mention that this general expression \eqref{fon1xp} clearly shows the
presence of the unitary operators $D$ and $S$ associated with $h(1)$ and $\su(1,1)$
respectively which is the contribution of the bosonic part of our SUSY
model. Moreover,
the fermionic contribution appears through the action of a unitary
operator associated with
$\su(2)$.

Now these states satisfy
the  MUR
\begin{equation}
( \Delta  X )^j_m  ( \Delta  P)^j_m  = \Delta^j_m
=
\frac{1}{2} \sqrt{1 + \frac{4 \delta^2 \sin^2 \phi}{{(1 - \delta^2 )}^2}}
\bigl|\langle  C
\rangle^j_m \bigr|.
\end{equation}

\begin{figure}
\centering
\begin{picture}(31.5,21)
\put(0,0){\framebox(31.5,21){}} \put(0,2.1){
\includegraphics[width=70mm]{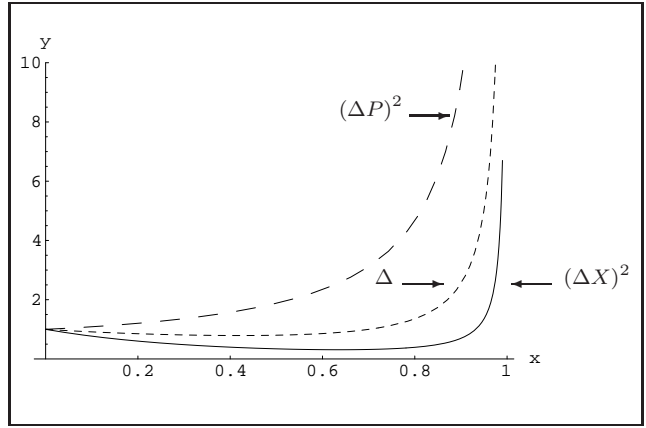}}
%\put(37,2){$\delta$}
\put(27.6,7){\scriptsize{${( \Delta  X)}^2$}}
 \put(27,7){\vector(-1,0){2}}
\put(16.4,15.4){\scriptsize{${( \Delta  P )}^2 $}}
\put(19.9,15.4){\vector(1,0){2}} \put(18.2,7){\scriptsize{$\Delta$}}
\put(19.6,7){\vector(1,0){2}}
\end{picture}
\caption{Graphs of the dispersions ${(\Delta X)}^2$, ${(\Delta P
)}^2$ and the factor $\Delta$ as functions of $x \equiv\delta $ for
$\phi= \pi / 6$, $|\tau|=|\mu|=1$, $j=1 / 2$.} \label{fig:vhuss19}
\end{figure}
The mean value of $ C$ is
\begin{equation}
\langle  C {\rangle}^j_m = {|\mu|}^2 + 2  {|\tau|}^2
\frac{{(1-\delta)}}{{(1
+
\delta)}}
\biggl(j - \frac{{ 4 ( j+ |m|) \delta }}{{{(1 + \delta)}^2} } \Omega
\biggr),
\end{equation}
where $\Omega$ is expressed in terms of Jacobi polynomials (see
Appendix~\ref{sec-su2ae}),
\begin{equation}
\Omega = \frac{ P^{(-2j,1)}_{j - |m| -1}  (1-  (8 \delta/(1 + \delta)^2))}
{P^{(-2j-1,0)}_{j - |m|} ( 1-  (8 \delta/(1 + \delta)^2))},
\end{equation}
for $ m= -j + 1, \dots, j-1$ and $\Omega=0$ for $m=\pm j$. In fact, we see
that in these
last cases, we have
\begin{equation}
\langle  C {\rangle}^j_{\pm j} = {|\mu|}^2 + 2  j   {|\tau|}^2
\frac{{(1-\delta)}}{{(1
+ \delta)}}. \label{extremosm}
\end{equation}

Its is now interesting to examine the behavior of the dispersions
$\Delta X$ and $\Delta  P$  in these states for the spin ${1/2}$
case. Using
 \eqref{moyab}
with  \eqref{extremosm}  for $j= {1/2}$, we get
\begin{eqnarray}
\nonumber ( {(\Delta  X)}^2 )_{\pm} =  \frac{(1- 2 \delta \cos\phi +
\delta^2)}{ 2 (1 -\delta^2)} \biggl[ {|\mu|}^2 +  {|\tau|}^2
\frac{{(1-\delta)}}{{(1 + \delta)}} \biggr],
\\
 ( {(\Delta  P)}^2 )_{\pm} = \frac{(1 + 2 \delta \cos\phi + \delta^2)}{
2 (1
-\delta^2)} \biggl[ {|\mu|}^2 +  {|\tau|}^2   \frac{{(1-\delta)}}{{(1 +
\delta)}} \biggr].
\end{eqnarray}
 with
\begin{equation}
\Delta_{\pm} = \frac{\sqrt{{(1-\delta^2)}^2 + 4 \delta^2 \sin^2\phi}}{ 2 (1-
\delta^2)}
\biggl[ {|\mu|}^2 +  {|\tau|}^2   \frac{{(1-\delta)}}{{(1 + \delta)}}
\biggr].
\end{equation}

If we take $\delta=0$  (i.e.\  $\lambda=1$) in these last
expressions, we find only the values of the dispersions of $X$ and
$P$ in the usual coherent states $| \psi \rangle^+ $ as given by
\eqref{junmedio}  and not the ones in the CS $|\psi \rangle^- $,
that is the reason why that  case has been treated separately.

\begin{figure}
\centering
\begin{picture}(31.5,21)
\put(0,0){\framebox(31.5,21){}} \put(0,2.1){
\includegraphics[width=70mm]{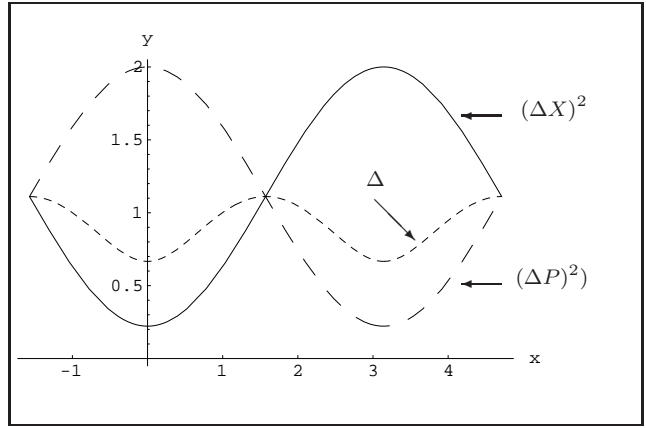}}
%\put(37,2){$\phi$}
\put(25.2,7){\scriptsize{${(\Delta P)}^2)$}}
\put(24.5,7){\vector(-1,0){2}} \put(25.2,15.4){\scriptsize{
${(\Delta X)}^2$}} \put(24.5,15.4){\vector(-1,0){2}}
\put(17.5,11.9){\scriptsize{ $\Delta$}}
\put(18.2,11.2){\vector(1,-1){2}}
\end{picture}
\caption{Graphs of the dispersions $ {(\Delta X)}^2 $, ${(\Delta
P)}^2$ and the factor $\Delta$ as functions of $x \equiv \phi$,
$\delta=0.5$, $|\tau|=|\mu|=1$, $j=1 / 2$.} \label{fig:vhuss15}
\end{figure}

Figures ~\ref{fig:vhuss19} and \ref{fig:vhuss15} show the behavior
of $\bigl( (\Delta X)^2 \bigr)_{\pm}$ and  $\bigl((\Delta P)^2
\bigr)_{\pm}$ and $\Delta_{\pm}$ as functions of $\delta $ for $\phi
={\pi/6}$ and as functions of $\phi$ for $\delta = 0.5$
respectively. We notice a similar behavior as for the position and
momentum operators.

\setcounter{equation}{0}\section{Construction of {\boldmath
$\lowercase{h}(1) \oplus \su(2)$} Hamiltonians} \label{sec-4uatro}

An application of our CS and SS based on the algebra $h(1) \oplus \su(2)$
will be the
study of possible
Hamiltonians which can be written as ${\mathcal H} = w {\mathcal A}^\dagger
{\mathcal
A}$, where ${\mathcal A}$ is a linear combination of the generators of $h(1)
\oplus
\su(2)$. It is clear that
the usual harmonic oscillator Hamiltonian will enter in the scheme as a
special case
($\S$\ref{sec-harmo}) but also the Jaynes--Cummings~\cite{kn:JaCu} one in
the
strong
coupling limit
($\S$\ref{sec-cano}) and ($\S$\ref{sec-nocano}).

Moreover, since the CS and SS already constructed in the preceding section
are
in fact
eigenstates of the operator $\mathcal A$, we
would be able to find easily some properties of the mean value and the
dispersion of the
associated  energies in those states.

\subsection{Isospectral {\boldmath $h(1) \oplus \su(2)$} harmonic oscillator
Hamiltonians}
\label{sec-harmo}

We are interested in systems for which the Hamiltonian is expressed in the
form

\begin{equation}
{\mathcal H} = w  {\mathcal A}^\dagger {\mathcal A}, \label{initialcalH}
\end{equation}
where
\begin{equation}
{\mathcal A} = \alpha_- a + \alpha_+ a^\dagger + \alpha_3 I + \beta_- J_+ +
\beta_+ J_- +
\beta_3 J_3,  \quad \alpha_- \ne 0,
\end{equation}
is an element of the $h(1) \oplus \su(2)$ algebra.  The commutator of the
operators
$\mathcal A$ and ${\mathcal A}^\dagger $ is
\begin{equation}
[ {\mathcal A}, {\mathcal A}^\dagger ]= \bigl({|\alpha_-|}^2 - {| \alpha_+
|}^2 \bigr) I
+ \bigl({|\beta_-|}^2 - {| \beta_+ |}^2 \bigr) J_3 +
 ( \beta_3 {\bar \beta}_+ - {\bar \beta}_3 \beta_-  ) J_+ +  (
{\bar
\beta}_3 \beta_+ - \beta_3 {\bar \beta}_- ) J_- .  \label{pri-commu}
\end{equation}
If $| Z \rangle$ is an eigenstate of the operator ${\mathcal A}$ with
eigenvalue $z$, i.e.
\begin{equation}
{\mathcal A} | Z \rangle = z | Z \rangle, \label{Zcohe}
\end{equation}
then the mean value of the energy in this state will always be given by
\begin{equation}
\langle Z | {\mathcal H} | Z \rangle = w {|z|}^2
\end{equation}
and the dispersion by
\begin{equation}
 (\Delta {\mathcal H})^2 =  w^2 |z|^2 \langle Z| [{\mathcal A}, {\mathcal
A}^\dagger ]| Z \rangle. \label{calhdisp}
\end{equation}

Firstly, let us consider the special case where
\begin{equation}
[ {\mathcal A}, {\mathcal A}^\dagger ]= I . \label{calhcomcano}
\end{equation}
This imposes the following conditions on the parameters:
\begin{equation}
{|\alpha_-|}^2 - {| \alpha_+ |}^2  = 1, \quad
|\beta_-| = | \beta_+ | \quad { \rm and} \quad
\beta_3 {\bar \beta}_+ - {\bar \beta}_3 \beta_- = 0,
\end{equation}
i.e.
\begin{equation}
\alpha_- = {\cosh\alpha}  e^{i \theta_-}, \quad \alpha_+ = {\sinh \alpha}
e^{i
\theta_+},\quad \beta_{\pm} = \beta e^{i \varphi_{\pm}}
\end{equation}
and
\begin{equation}
\beta_{3} = \begin{cases}  r e^{i (\varphi_+ + \varphi_-) /2},\ r
\in {\mathbb R}_+ \cup \{ 0 \} &   if
 \beta \ne 0 \cr
r e^{i \varphi_3}, \ r \in {\mathbb R}_+ \cup \{0 \}  & if  \beta =
0 . \cr \end{cases}
\end{equation}

When $\beta \ne 0$, the operator $\mathcal A$ then takes the form
\begin{equation}
{\mathcal A} = \cosh\alpha   e^{i \theta_-}   a +  \sinh\alpha   e^{i
\theta_+}
a^\dagger + \alpha_3 I +
\beta  ( e^{i \varphi_-} J_+ + e^{i \varphi_+} J_-  ) + r e^{i
(\varphi_+ +
\varphi_-) /2} J_3.
\end{equation}
The  parameter $b$ given in \eqref{b-factor} becomes $b= \sqrt{4 \beta^2 +
r^2}  e^{i
(\varphi_+ + \varphi_-) / 2} $ and is different from zero.
Therefore in this case, according to the equation \eqref{fon1}, the
normalized
solutions of the eigenstates equation \eqref{Zcohe} are given
by
\begin{equation}
|Z \rangle^j_m = S(\Lambda) D \bigl(\zeta_m (\alpha_3,1)\bigr) T D ( z e^{
-i
\theta_-})
|0; j , m \rangle,  \label{sdtzcohe}
\end{equation}
where
\begin{equation}
\Lambda= - \alpha e^{i (\theta_+ - \theta_-)}, \quad
\zeta_m (\alpha_3, \epsilon) = - [ \alpha_3 + \epsilon  m   \sqrt{4
\beta^2 + r^2}
e^{i (\varphi_+ + \varphi_-) / 2}] e^{-i \theta_-}
\end{equation}
and
\begin{equation}
T = \exp\biggl( - \frac{{\tilde \theta}}{2} [ e^{-i (\varphi_+ -
\varphi_-) / 2} J_+
- e^{i (\varphi_+ - \varphi_-) / 2} J_-  ] \biggr),
\end{equation}
with
\begin{equation}
\frac{{\tilde \theta}}{2} = \tan^{-1} \biggl(
\sqrt{1 - \frac{r}{2\beta^2} ( \sqrt{4 \beta^2 + r^2} - r )} \biggr).
\end{equation}
This means that T is an unitary operator.

We remark that, if we define the new operator
\begin{eqnarray}
\nonumber
{\mathcal A}_0
        &=& D^\dagger (- \alpha_3 e^{-i \theta_-}) S^\dagger (\Lambda)
{\mathcal A} S(\Lambda) D(- \alpha_3 e^{-i \theta_-})  \\
        &=& e^{i \theta_-} a +
\beta  ( e^{i \varphi_-} J_+ + e^{i \varphi_+} J_- ) + r e^{i
(\varphi_+ + \varphi_-) /2} J_3,  \label{opcalA0}
\end{eqnarray}
which is simpler than the original ${\mathcal A}$, then the new Hamiltonian
${\mathcal
H}_0 = w {\mathcal A}_0^\dagger {\mathcal A}_0$ is isospectral to the
Hamiltonian $\mathcal H \equiv$ \eqref{initialcalH}.

The dispersion of $\mathcal H$ calculated on the states \eqref{sdtzcohe} is,
from  \eqref{calhdisp}  and  \eqref{calhcomcano}, given by
$(\Delta {\mathcal H})^2 = w^2 {|z|}^2 $ and is the same as  the
one of
${\mathcal H}_0 $ calculated
on the states    $D\bigl(\zeta_m (-z,1)\bigr) T |0 ; j,m \rangle $. This
value
is exactly
the dispersion of the harmonic oscillator in the usual CS.

On the other hand, due to  \eqref{calhcomcano}  we have $[ {\mathcal H},
{\mathcal A}] = - w {\mathcal A}$, so we have a complete analogy with the
harmonic
oscillator.
The CS associated to the Hamiltonian
${\mathcal H}$, called generalized harmonic oscillator, are those given by
the
equation  \eqref{sdtzcohe}  and  thus, one can write them in the form
\begin{equation}
| Z \rangle^j_m =  {\mathcal D}(z)  |{\tilde 0} \rangle^j_m, \text{ where }
{\mathcal
D}(z) = \exp(z {\mathcal A}^\dagger - \bar z {\mathcal A})
\end{equation}
and $|{\tilde 0} \rangle^j_m$, $m=-j, \dots, j$,  are the fundamental
states
of the
system ${\mathcal H}$, that is the eigenstates of ${\mathcal H}$
corresponding
to the
$(2j +1)$ degenerate eigenvalue 0. They are also eigenstates of $\mathcal A$
corresponding to the eigenvalue 0. So they can be written
\begin{equation}
|{\tilde 0} \rangle^j_m = S(\Lambda) D \bigl(\zeta_m(\alpha_3,1)\bigr) T |0;
j
, m
\rangle. \label{estado-base}
\end{equation}

Furthermore, the $SS$ associated with ${\mathcal H}$, are
given by
\begin{equation}
|\tilde \psi \rangle^j_m = {\mathcal S}(\chi) {\mathcal D}(z) |\tilde 0
\rangle_m^j,
\end{equation}
where the supersqueezed operator ${\mathcal S}(\chi)$
is given by $\exp({\chi} {{\mathcal A}^{\dagger}}^2/2 - \overline{\chi}
{\mathcal A}^2/2)$ and the
superdisplacement operator ${\mathcal D}(z)$ is given in (4.17). If we
define ${\mathcal X} = ({\mathcal A} + {\mathcal A}^\dagger)/\sqrt{2}$ and
${\mathcal P}
= i({\mathcal A}^\dagger - {\mathcal A})/\sqrt{2}$, these states (4.19) minimize
the
SRUR $(\Delta {\mathcal X})^2 (\Delta {\mathcal P})^2 = \bigl( 1+\langle F
\rangle^2\bigr)/4$, i.e.\ they are solutions of the eigenstate equation
$\bigl[(1-\lambda){\mathcal A}^\dagger + (1+\lambda){\mathcal A}\bigr]
|\psi\rangle = \sqrt{2} |\psi \rangle$.

The eigenstates of $\mathcal H$ corresponding to the $(2j+1)$ degenerate
energy
eigenvalue $E_n = n w$ are now given by
\begin{equation}
|{\tilde n} \rangle^j_m = \frac{{{\mathcal A}^\dagger}^n}{\sqrt{n!}}
|{\tilde 0} \rangle^j_m.   \label{tilde-vacua}
 \end{equation}
These states may be obtained as the action of an unitary operator on the
states
$|n; j, m \rangle$. Indeed, if we introduce the unitary operator
\begin{equation}
U_n^m = e^{- i n \theta_-} S(\Lambda) D \bigl(\zeta_m (\alpha_3,1)\bigr) T,
\end{equation}
we see that, from  \eqref{tilde-vacua}, we have
\begin{eqnarray}
|{\tilde n} \rangle^j_m
        &=& \frac{e^{i n \theta_-}}{\sqrt{n !}} {(
{\mathcal
A}^\dagger )}^n U_n^m |0; j,m \rangle, \nonumber \\
        &=& \frac{e^{i n \theta_-}}{\sqrt{n !}} U_n^m {\bigl( {( U_n^m
)}^\dagger
{\mathcal
A}^\dagger U_n^m \bigr)}^n |0;j,m \rangle, \nonumber \\
        &=& \frac{e^{i n \theta_-}}{\sqrt{n !}} U_n^m  {\bigl( e^{- i
\theta_-}
a^\dagger +
\sqrt{4 \beta^2 + r^2} e^{- i (\varphi_+ + \varphi_-) /2}
 (J_3 - m) \bigr)}^n |0;j,m \rangle.
\end{eqnarray}
Since we have $(J_3 - m) |0; j,m \rangle = 0$, we finally find
\begin{equation}
|{\tilde n} \rangle^j_m = U^m_n |n ;j,m \rangle.
\end{equation}

In the case $\beta =0$, the operator ${\mathcal A}$ is given by
\begin{equation}
{\mathcal A} = \cosh\alpha    e^{i \theta_-}   a +  \sinh\alpha    e^{i
\theta_+}
a^\dagger + \alpha_3 I + r e^{i \varphi_3} J_3.
\end{equation}
Then,  if $r \ne 0$,  one has the same results  as above, except that it is
necessary to
replace T by I and
$b$ by $\beta_3 = r e^{i \varphi_3}$. If $r=0$, ${\mathcal A}$ is an element
of the
algebra $h(1)$ and then the results are the  ones obtained in
Section~\ref{sec-uno}
for the standard harmonic oscillator after applying the unitary
transformation
$S(\Lambda) D(- \alpha_3 e^{- i \theta_- })$.

\subsection{Strong-coupling limit of the Jaynes--Cummings Hamiltonian as
limit
of $h(1) \oplus \su(2)$ Hamiltonians}
\label{sec-cano}

We are going to consider now the case where
\begin{equation}
[ {\mathcal A} , {\mathcal A}^\dagger ] = I + 2 x J_3, \quad x \in {\mathbb
R}.
\label{com-calcano}
\end{equation}
This imposes the following conditions on the parameters:
\begin{equation}
{|\alpha_-|}^2 - {| \alpha_+ |}^2  = 1, \quad
{|\beta_-|}^2  - {| \beta_+ |}^2 = x  \quad { \rm and} \quad
\beta_3 {\bar \beta}_+ - {\bar \beta}_3 \beta_- = 0. \label{xconditions}
\end{equation}
We already know the results when $x=0$. When $x \ne 0$, the conditions
 \eqref{xconditions}  imply
\begin{equation}
\alpha_- = {\cosh\alpha}  e^{i \theta_-}, \quad \alpha_+ = {\sinh \alpha}
   e^{i \theta_+},
\quad \beta_{3} = 0
\end{equation}
and
\begin{equation}
\beta_- = \begin{cases} x^{1 / 2} \cosh\beta  e^{i \varphi_-}, &
{\rm if} \
 x
> 0   \cr {|x|}^{1 /2} \sinh\beta    e^{i \varphi_-}, & {\rm if} \  x < 0 ,
\cr \end{cases}
\end{equation}
\begin{equation}
\beta_+ = \begin{cases}  x^{1 / 2} \sinh\beta   e^{i \varphi_+}, &
i{\rm if} \  x
> 0  \cr {|x|}^{1 /2} \cosh\beta    e^{i \varphi_+},   &  {\rm if} \ x <
0 . \cr \end{cases}
\end{equation}
The parameter $b \equiv$ \eqref{b-factor}  becomes  $b= |x|^{1/2} \sqrt{2
\sinh(2 \beta)}
e^{i (\varphi_+ + \varphi_-) /2}$,
this means that $b=0$  if and only if $\beta=0$.

In the case $\beta \ne 0$, according to the equations  \eqref{fon1},
 \eqref{opt1},
 \eqref{eifi}, and  \eqref{themedio}, the normalized eigenstates of the
operator $\mathcal A$ are given by \beqa \nonumber | Z (x)
\rangle^j_m
        &=& {\bigl(C^j_m (x)\bigr)}^{- 1 / 2} S(\Lambda) D(-
\alpha_3 e^{-i \theta_-}) D(\eta_m (z,x)) \\
        &&  \times \exp\biggl[ -
\frac{x}{2|x|} \ln (\tanh \beta  ) J_3 \biggr] U | 0; j,m \rangle,
\label{quasisuper} \eeqa where
\begin{equation}
\eta_m (z,x) = \bigl[ z - m |x|^{1 /2} \sqrt{2 \sinh(2 \beta)} e^{i
(\varphi_+
+
\varphi_-) /2} \bigr] e^{-i \theta_-},
\end{equation}
\begin{equation}
U = \exp \biggl[- \frac{\pi}{4} \bigl( e^{-i (\varphi_+ - \varphi_-) / 2}
J_+ -
e^{i
(\varphi_+ - \varphi_-) / 2} J_- \bigr) \biggr]
\end{equation}
and
\begin{eqnarray}
\nonumber
C^j_m (x)
        &=& \langle j , m | U^\dagger  \exp\biggl[ - \frac{x}{|x|}
\ln (\tanh \beta
 ) J_3 \biggr] U | j,m \rangle  \\
        &=& {\biggl( \frac{1 + \tanh \beta}{2
\sqrt{\tanh\beta}}\biggr)}^{\mp 2
m}
P^{0;
\mp 2
m}_{j \pm m} \biggl( \frac{ 1 + \tanh^2 \beta}{2 \tanh\beta}\biggr).
\end{eqnarray}
\begin{figure} \centering
\begin{picture}(31.5,21)
\put(0,0){\framebox(31.5,21){}}
%\multiput(0,0)(0,5){7}{\line(1,0){45}}
%\multiput(0,0)(5,0){10}{\line(0,1){30}}
\put(1.83,5.5){{\scriptsize{$|x|=1$}}}
\put(2.45,9.8){{\scriptsize{$|x|=2$}}}
\put(3.5,12.6){{\scriptsize{$|x|=4$}}} \put(28.35,2.45){\scriptsize{
$\beta$}} \put(27.65,4.91){\scriptsize{ $w^2 {|z|}^2$}}
\put(0.91,19.32){\scriptsize{ ${( \Delta {\cal H})}^2_{\pm} $}}
\put(1.05,1.4){\includegraphics[width=70mm]{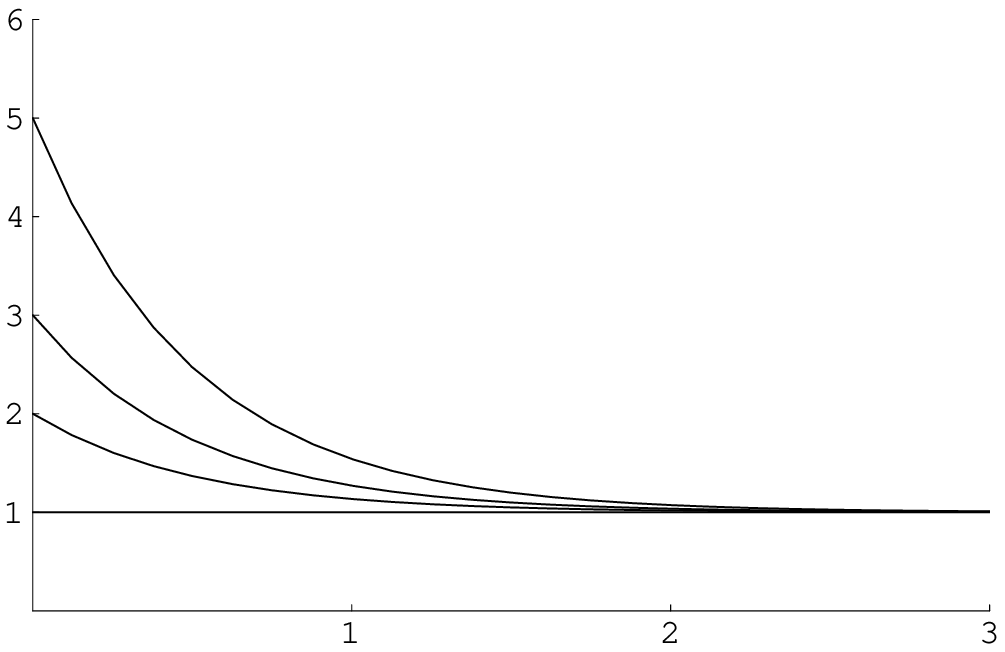}}
\end{picture}
\caption{Graphs of the dispersions ${ \left( {( \Delta {\cal H} )
}^2 \right)}_{\pm} \equiv$(\ref{exp-2beta}) as functions of $\beta >
0$ for $|x|=0, 1, 2, 4$.} \label{fig:dispersion1}
\end{figure}
From  \eqref{calhdisp}  and  \eqref{com-calcano}, the dispersion of the
${\mathcal H}
\equiv$  \eqref{initialcalH}  in the states  \eqref{quasisuper}  can be
calculated
explicitly. We get
\begin{equation}
{ \bigl{( \Delta {\mathcal H} ) }^2 \bigr)}^j_m = w^2 |z|^2 \bigl( 1  +  2
x  {}^j_m
\bigl\langle Z (x) | J_3 | Z (x) \bigr\rangle^j_m  \bigr).
\end{equation}
In the last expression, the mean value of $J_3$ is obtained  in a similar
way
then to get \eqref{moyj3f}. The result is
\begin{equation}
{}^j_m \bigl\langle Z (x)| J_3 | Z (x) \bigr\rangle^j_m = \frac{x}{|x|}
\biggl\{
|m|
e^{-
2 \beta} +
\frac{(j + |m| + 1)}{2 \sinh (2\beta)}
\frac{P^{1; 2| m|}_{j - |m| -1}  ( \coth (2 \beta)  )}{P^{0 ; 2|
m|}_{j - |m|} ( \coth (2 \beta) ) } \biggr\}.
\end{equation}
If we take $m = \pm j$, the dispersion of $\mathcal H$ is
\begin{equation}
{ \bigl( {( \Delta {\mathcal H} ) }^2 \bigr)}^j_{\pm j} = w^2 |z|^2 \bigl( 1
+  2  j
|x|  e^{- 2 \beta}  \bigr)
\end{equation}
and, in particular, when $j = { 1/2}$, we get
\begin{equation}
{ \bigl( {( \Delta {\mathcal H} ) }^2 \bigr)}_{\pm} = w^2 |z|^2 \bigl( 1  +
|x|  e^{- 2
\beta}  \bigr). \label{exp-2beta}
\end{equation}
Fig.~\ref{fig:dispersion1} shows the graphs of ${\bigl( {(\Delta {\mathcal
H})}^2
\bigr)}_{\pm}$ as functions of $\beta$ for different values of $|x|$ when
$w^2 {|z|}^2 $ is taken equal to $1$.

Let us compute the new operator ${\mathcal A}_0$ defined as
\eqref{opcalA0}.
We get
\begin{equation}
{\mathcal A}_0 = \begin{cases}e^{i \theta_-}   a  + x^{1 /2}
\cosh\beta e^{i \varphi_-} J_+ +  x^{1 /2} \sinh\beta   e^{i
\varphi_+} J_-, & {\rm if} \ x > 0   \cr e^{i \theta_-}   a  +
{|x|}^{1 /2} \sinh\beta e^{i \varphi_-} J_+  + {|x|}^{1 /2}
\cosh\beta   e^{i \varphi_+} J_-, &   {\rm if} \ x < 0
\end{cases},
\end{equation}
and  a new Hamiltonian ${\mathcal H}_0 = w {\mathcal A}_0^\dagger {\mathcal
A}_0$
isospectral to the Hamiltonian
$\mathcal H$ which takes the form
\begin{eqnarray}
{\mathcal H}_0
        &=&  w \bigl\{  a^\dagger a + |x| \bigl[ \sinh^2 (\beta) J_- J_+
+ \cosh^2 (\beta)  J_+ J_-   \bigr]\nonumber \\ && + {|x|}^{1/ 2}
\cosh\beta [ e^{i (\varphi_+ - \theta_-)} a^\dagger J_- + e^{- i
(\varphi_+ - \theta_-)} a J_+] \nonumber
  \\
        && + {|x|}^{1/ 2} \sinh\beta [ e^{i (\varphi_- - \theta_-)}
a^\dagger
J_+ + e^{- i (\varphi_- - \theta_-)} a J_-] \nonumber\\
            && + |x| \sinh\beta \cosh\beta [ e^{i (\varphi_+ -
\varphi_-)}
J_-^2 +  e^{- i (\varphi_+ - \varphi_-)} J_+^2 ] \bigr\}, \label{H0x}
\end{eqnarray}
if $x < 0$. If  $x > 0$, we get a similar expression except that we must
make
the change $\sinh\beta \leftrightarrow \cosh\beta$.

In the spin-${1/2}$ representation, we have
\begin{equation}
J_-^2 = J_+^2=0, \quad J_+ J_- = \frac{I}{2} + J_3 \quad {\rm and} \quad
J_- J_+ = \frac{I}{ 2} - J_3,
\end{equation}
hence  \eqref{H0x}  becomes \beqa \nonumber {\mathcal H}_0
        &=& w \biggl\{  \biggl( a^\dagger a + \frac{I}{2}\biggr) - x
J_3  + {|x|}^{1/ 2} \cosh\beta [ e^{i (\varphi_+ - \theta_-)} a^\dagger J_-
+
e^{- i
(\varphi_+ - \theta_-)} a J_+]   \\
        && +    {|x|}^{1/ 2} \sinh\beta  [ e^{i (\varphi_-
- \theta_-)} a^\dagger J_+ + e^{- i (\varphi_- - \theta_-)} a J_-] +
\bigl( |x| \cosh (2 \beta) - 1 \bigr) \frac{I }{2} \biggr\}
\label{quasi-cumming} \eeqa and a similar expression when $x >0$,
making the literal change  $\sinh\beta \leftrightarrow \cosh\beta$.
If we take $x = - {w_0/w}$, $\varphi_+ = \theta_- $  and the limit
$\beta \mapsto 0$, then ${\mathcal H}_0 \equiv$
\eqref{quasi-cumming}  becomes
\begin{equation}
{\mathcal H}_0 = w \biggl( a^\dagger a + \frac{1}{2} \biggr) + w_0 J_3 +
\sqrt{w
w_0}(a^\dagger J_- + a J_+) + \frac{w - w_0}{2} I, \label{jc-up-const}
\end{equation}
which is the Jaynes--Cummings Hamiltonian~\cite{kn:JaCu} up to a constant
term
and for a coupling constant given by $\kappa = \sqrt{ww_0}$. Let us recall
that
this
Hamiltonian describes the interaction of a cavity
mode (with frequency $w$) with a two level-system ($w_0$ being the atomic
frequency).
When $x=-1$, i.e., for  $w=w_0$,  \eqref{jc-up-const}  becomes the
strong-coupling limit
of the Jaynes--Cummings Hamiltonian.

In the case $\beta =0$, the new operator ${\mathcal A}_0
\equiv$ \eqref{opcalA0} reduces
now to
\begin{equation}
{\mathcal A}_0 (x) = \begin{cases} e^{i \theta_-} a + {|x|}^{1 /2}
e^{i \varphi_+} J_-, & {\rm if} \ x < 0 \cr
 e^{i \theta_-} a + {|x|}^{1 /2} e^{i \varphi_-} J_+, & {\rm if} \ x > 0  \cr \end{cases}.
\label{non-canA0}
\end{equation}
As we have here $b=0$, according to the expressions \eqref{solb++b3=0} and
 \eqref{solb--b3=0}, the
orthonormalized eigenstates of ${\mathcal A}_0$ are given by \beqa |
Z (x) \rangle^j_m
         =  {\bigl({\widetilde C}^j_m (x)\bigr)}^{1 /2} D (z e^{-i\theta_-
}) \qquad \qquad \qquad  \qquad \qquad \qquad \qquad \qquad \qquad \qquad \nonumber \\
\times  \sum_{k=0}^{j-m} {(-1)}^k {j-m  \choose k}
\frac{(2j-k)!}{(2j)!} { (e^{-i \theta_- } a^\dagger)}^{j-m-k}
{\biggl( J_{\mp} \frac{e^{-i \varphi_{\mp}}}{\sqrt{|x|}}\biggr)}^k
\biggl| 0; j , \frac{x}{|x|} j \biggr\rangle,\label{zsuper} \eeqa
where the $-$ sign refers to $ x > 0$ and the sign $+$ to $x < 0 $
and
\begin{equation}
 {\widetilde C}^j_m (x) = (j-m)! \sum_{k=0}^{j-m} {j-m \choose k} \frac{(2j
-k)!}{
(2j)!}
{\biggl( \frac{1}{|x|}\biggr)}^k. \label{fac-nor-supco}
\end{equation}

Since in this case, we have
\begin{equation}
{}^j_m \bigl\langle  Z (x)\bigr| J_3 \bigl| Z (x) \bigr\rangle^j_m  =
\frac{x}{
|x|}
\biggl[ j + |x| \frac{\partial}{\partial |x|}  \ln \bigl({\widetilde C}^j_m
(x)\bigr)
\biggl],
\end{equation}
the dispersion of ${\mathcal H}_0 = w {\mathcal A}_0^\dagger {\mathcal A}_0
$
in the
states  \eqref{zsuper}  is given by
\begin{equation}
{ \bigl( {( \Delta {\mathcal H}_0 ) }^2 \bigr)}^j_{m} = w^2 {|z|}^2  \biggl[
1
+ 2 |x| j + 2 |x|^2 \frac{\partial}{\partial |x|}  \ln \bigl({\widetilde
C}^j_m
(x)\bigr) \biggr].
\end{equation}
When $m = j$, we have $ {\widetilde C}^j_m (x) =1$, so that we get
\begin{equation}
{ \bigl( {( \Delta {\mathcal H}_0 ) }^2 \bigr)}^j_{j} =  w^2 {|z|}^2  \bigl(
1 +
2
|x| j \bigr).
\end{equation}

\begin{figure}\centering
\begin{picture}(31.5,21)
\put(0,0){\framebox(31.5,21){}}
%\multiput(0,0)(0,5){7}{\line(1,0){45}}
%\multiput(0,0)(5,0){10}{\line(0,1){30}}
\put(0.7,1.4){\includegraphics[width=70mm]{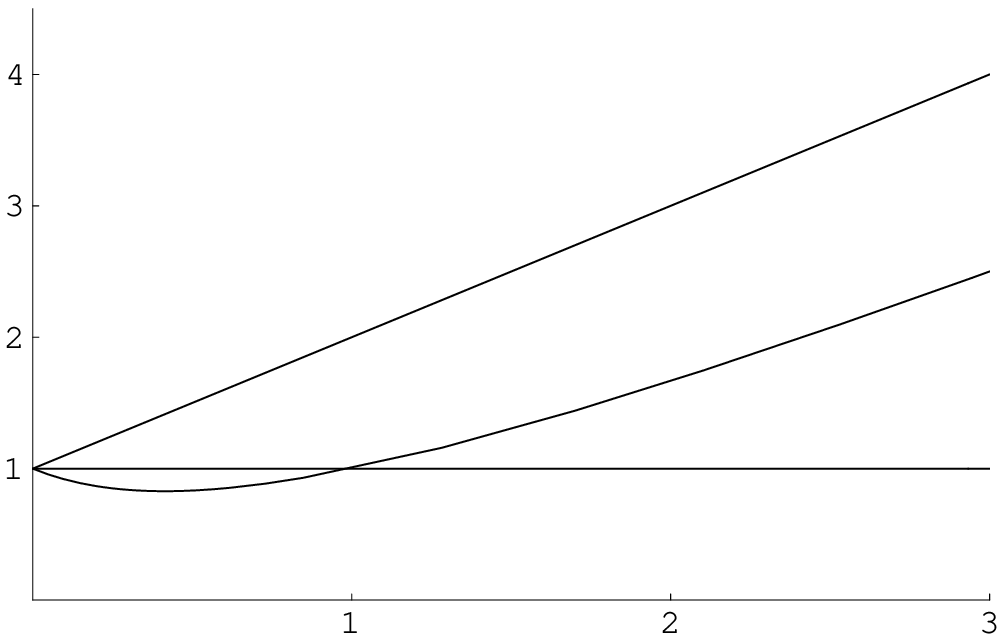}}
\put(28,2.45){\scriptsize{ $|x|$}} \put(27.3,5.95){\scriptsize{ $
w^2 {|z|}^2 $ }} \put(27,11.2){\scriptsize{ ${( \Delta {\cal H}_0
)}^2_- $}} \put(27,16.8){\scriptsize{ ${( \Delta {\cal H}_0 )}^2_+
$}}
\end{picture}
\caption{Graphs of the dispersions ${ \left( {( \Delta {\cal H}_0 )
}^2 \right)}_{\pm}$ as given by (\ref{x-linear}) and (\ref{x-curve})
as functions of $|x|$.} \label{fig:xminimum}
\end{figure}

For example,  when $j= {1/2}$, the dispersion corresponding to $m=
{1/2}$ is
given by
\begin{equation}
{ \bigl( {( \Delta {\mathcal H}_0 ) }^2 \bigr)}_+ = w^2 {|z|}^2  \bigl( 1 +
|x|
\bigr)
\label{x-linear}
\end{equation}
and one obtains the same result as in the preceding case when we
take  the limit $\beta \mapsto 0$. On the other hand, for
$m=-{1/2}$,  we get
\begin{equation}
{ \bigl( {( \Delta {\mathcal H}_0 ) }^2\bigr)}_- = w^2 {|z|}^2 \biggl[ 1 +
|x| \frac{(|x| - 1)}{(|x| + 1) }\biggr] \label{x-curve}
\end{equation}
and it is always smaller than ${ \bigl( {( \Delta {\mathcal H}_0 )
}^2 \bigr)}_+ $. In this last case, we see that if $|x| > 1$, the
dispersion is bigger than $w^2 {|z|}^2$ while  if $|x| <1$ it is
smaller than $w^2 {|z|}^2$ and if $|x|=1$ it is equal to $w^2
{|z|}^2$. Furthermore, the dispersion reaches its minimum $0.83 w^2
{|z|}^2$ when $|x|= (\sqrt{2} -1)$. Fig.~\ref{fig:xminimum} shows
the behavior of dispersions ${ \bigl( {( \Delta {\mathcal H}_0 ) }^2
\bigr)}_{\pm}$ as function of $|x|$.

Let us finally mention that the Hamiltonian ${\mathcal H}_0$ in this case
and
for $j = {1/2}$ corresponds to \eqref{quasi-cumming} when $\beta = 0$. A
special
case is again the Jaynes--Cummings Hamiltonian \eqref{jc-up-const} so we get
eigenstates of ${\mathcal A}_0 \equiv$ \eqref{non-canA0}  such that the
dispersion
of this Hamiltonian is minimized and lower than $w^2 {|z|}^2$.

\subsection{Generalized {\boldmath $h(1) \oplus \su(2)$} non-canonical commutation
relation}\label{sec-nocano}

In the case where we have
\begin{equation}
[{\mathcal A} , {\mathcal A}^\dagger] = I + \gamma J_+ + {\bar \gamma} J_-,
\quad \gamma \in {\mathbb C},\ \gamma \ne 0,
\label{relnotcano}
\end{equation}
according to  \eqref{pri-commu}, the necessary conditions on the original
parameters are
\begin{equation}
{|\alpha_-|}^2 - {|\alpha|_+}^2 =1, \quad |\beta_-| = \beta_+, \quad \beta_3
{\bar \beta}_+ - {\bar \beta}_3 \beta_- = \gamma = \rho e^{i  \nu},
\end{equation}
where $\rho \in {\mathbb R}_+$. A suitable choice of the parameters is
\begin{equation}
\alpha_- = \cosh\alpha  e^{i \theta_-}, \quad \alpha_+ = \sinh\alpha  e^{i
\theta+}, \quad \beta_{\pm} = \beta e^{i \varphi_{\pm}}, \quad \beta_3 = r
e^{i
\varphi_3}, \quad \beta \ne 0, \ r \ne 0,
\end{equation}
such that
\begin{equation}
r \beta [ e^{i (\varphi_3 - \varphi_+)} - e^{-i(\varphi_3 - \varphi_+)} ] =
\rho e^{i \nu}.  \label{roinu}
\end{equation}
Equation \eqref{roinu} implies that
\begin{equation}
\rho = 2 r \beta \biggl| \sin \biggl( \varphi_3 - \frac{(\varphi_+ +
\varphi_-)}{2
} \biggr) \biggr|
\end{equation}
and the following conditions on the phases:
$\varphi_3 \ne {(\varphi_+ + \varphi_-)/2}$, $ \varphi_3 \ne {(\varphi_+
+ \varphi_-)/2} + \pi$ and$\varphi_+ - \varphi_- = \pi - 2 \nu,  \nu \in [0
,3
{\pi/2}]$ or  $\varphi_+ - \varphi_- = 3 \pi - 2 \nu$,  $\nu \in [{\pi/2}, 2
\pi]$.
Thus, the
operator
${\mathcal A}$ compatible with all the previous conditions is
\begin{equation}
{\mathcal A} = \cosh\alpha  e^{i \theta_-} a +  \sinh\alpha  e^{i \theta_+}
a^\dagger + \alpha_3 I +
e^{i (\varphi_- - \nu)} \biggl[\beta  ( e^{ i \nu} J_+  - e^{- i \nu} J_-
) +
\frac{\rho}{2 \beta |\cos\theta|} e^{i \theta} J_3 \biggr],
\end{equation}
where
\begin{equation}
\theta = \varphi_3 - (\varphi_- - \nu), \quad  - \frac{\pi}{2} < \theta < 3
\frac{\pi}{2}.
\end{equation}
The new operator ${\mathcal A}_0$ defined in  \eqref{opcalA0}  is then given
by
\begin{equation}
{\mathcal A}_0 = e^{i \theta_-} a + e^{i (\varphi_- - \nu)} \biggl[
\beta ( e^{ i \nu} J_+  - e^{- i \nu} J_- ) + \frac{\rho}{ 2 \beta
|\cos\theta|} e^{i \theta} J_3 \biggr].
\end{equation}
The parameter $b\equiv$ \eqref{b-factor} is now   $b = {i\sqrt{16 \beta^2 \cos^2
(\theta) -  \rho^2 e^{2 i \theta} } e^{i (\varphi_-  - \nu) }/(2 \beta
|\cos\theta|)} $, i.e.\ $b = 0$ if and only if $\beta =
{\sqrt{\rho}/2}$
and
$\theta = \pi$.

Here we can proceed as before, that is, when $b=0$, find, by means of the
equation~\eqref{solb+b-b3=0} the eigenstates
of ${\mathcal A}_0$ and, when $b \ne 0$, find the solutions by means of the
equation \eqref{fon1} and then calculate the dispersions of ${\mathcal
H}_0$.

But, we will follow another treatment which teachs us  about
the similarities between the canonical and the non-canonical cases. Indeed,
seen in another perspective, the commutation relation \eqref{relnotcano} can
be
expressed in the form
\begin{equation}
[{\mathcal A}_0, {\mathcal A}^\dagger_0 ] = I + 2 \rho {\mathbb J}_3,
\end{equation}
where we have set
\begin{equation}
{\mathbb J}_3 = \frac{(e^{i \nu} J_+ +  e^{- i \nu} J_- )}{2}.
\end{equation}
Thus, when  $b=0$,  ${\mathcal A}_0$  becomes
\begin{equation}
{\mathcal A}_0 = e^{i \theta_-} a + \sqrt{\rho} e^{i ( \varphi_- -  \nu)}
{\mathbb J}_+ ,
\end{equation}
with
\begin{equation}
{\mathbb J}_{\pm}  = \pm \frac{(e^{i \nu} J_+ -  e^{- i \nu} J_- )}{
2} - J_3.
\end{equation}

The operators ${\mathbb J}_3$, ${\mathbb J}_{\pm}$ satisfy the $\su(2)$
algebra
and let us denote by $|J,M\rangle$ the eigenstates of both ${\mathbb J}^2$ and
${\mathbb J}_3$. We have again:
\begin{equation}
{\mathbb J}_3 | J , M \rangle = M | J , M \rangle, \quad
{\mathbb J}_{\pm}  | J , M \rangle = \sqrt{ (J \mp M) ( J \pm M + 1)}  | J ,
M
\rangle.
\end{equation}

Now, it is clear that the resolution of the problem to find the
eigenstates of ${\mathcal A}_0$ is similar to the canonical case.
Indeed, the normalized eigenstates of ${\mathcal A}_0$ are given by
\beqa | Z (\rho) \rangle^J_M
        = {\bigl({\widetilde C}^J_M (\rho) \bigr)}^{1 /2} D (z e^{-i
\theta_-}) \qquad \qquad \qquad \qquad  \qquad \qquad \qquad \qquad \qquad \qquad \nonumber\\
\label{mathbbzsuper}
  \times      \sum_{k=0}^{J-M} {(-1)}^k {J-M  \choose
k} \frac{(2J-k)!}{(2J)!} { ( e^{-i \theta_-} a^\dagger
)}^{J-M-k}{\biggl( \frac{{\mathbb J}_{-}  e^{-i (\varphi_ - -
\nu)}}{ \sqrt{\rho}}\biggr)}^k  | 0 ; J ,  J \rangle, \eeqa where
${\widetilde C}^J_M (\rho)$ is given as in  \eqref{fac-nor-supco}.

As before, the dispersion of ${\mathcal H}_0$ in the states
\eqref{mathbbzsuper}
 is
given by
\begin{equation}
\bigl( (\Delta {\mathcal H}_0)^2 \bigr)^J_M = w^2 {|z|}^2 \biggl[ 1 + 2 J
\rho + 2 \rho^2\frac{\partial}{\partial \rho} \ln \bigl(  {\widetilde C}^J_M
(\rho) \bigr)  \biggr].
\end{equation}
For example, when $J={1/2}$, we have
\begin{equation}
\bigl( (\Delta {\mathcal H}_0)^2 \bigr)_+ = w^2 {|z|}^2  ( 1 + \rho),
\quad
\bigl( (\Delta {\mathcal H}_0)^2 \bigr)_- = w^2 {|z|}^2  \biggl[ 1 + \rho
\frac{(\rho -1)}{(\rho + 1)} \biggr].
\end{equation}
Evidently, the behavior of these dispersions as functions of $\rho$
is identical to that described in the last paragraph of the previous
section.

In the general case where  $b \ne 0$, ${\mathcal A}_0$ can be expressed in
the
form
\begin{equation}
{\mathcal A}_0 = e^{i \theta_-} a + e^{i (\varphi_- - \nu)}
\biggl\{\biggl[\frac{4 {\beta^2} {|\cos\theta|} - {\rho} {e^{i \theta}}}{4
{\beta
|\cos\theta|}}\biggr] {\mathbb J}_+  -\biggl[\frac{4 {\beta^2} {|\cos\theta|}
+ {\rho e^{i\theta}}}{4 {\beta |\cos\theta|}}\biggr] {\mathbb J}_- \biggr\}.
\end{equation}
From  \eqref{fon1}, we see that the eigenstates of ${\mathcal A}_0$ are
\begin{equation}
|Z \rangle^J_M = ( C^j_m)^{-1 /2} D( z e^{- i \theta_-}) T_{\mathrm{eff}} |
0 ;
J , M
\rangle,
\end{equation}
where
\begin{equation}
T_{\mathrm{eff}} = e^{\Phi_- {\mathbb J}_+ }   e^{\Phi_+ {\mathbb J}_- },
\end{equation}
with
\begin{equation}
\Phi_-  = i \frac{[4 \beta^2 |\cos\theta| - \rho e^{i\theta}]}{R^{1/2}
e^{i\tilde \varphi /2}}, \quad
\Phi_+  = i \frac{[4 \beta^2 |\cos\theta| + \rho e^{i \theta}]}
{2 R^{1 /2} e^{i\tilde\varphi} /2}.
\end{equation}

The dispersion of ${\mathcal H}_0$ in these states is
\begin{equation}
\bigl( (\Delta {\mathcal H}_0)^2 \bigr)^J_M = w^2|z|^2  \bigl[  1 + 2 \rho
{}^J_M \langle Z |{\mathbb J}_3 |Z \rangle^J_M  \bigr],
\end{equation}
where \cite{kn:Nam}
\begin{equation}
{}^J_M \langle Z | {\mathbb J}_3 | Z \rangle^J_M =  M
\biggl( \frac{ 1 - {|\Phi_- |}^2}{1 +  {|\Phi_- |}^2 } \biggr)  +
\frac{(J-M+1)}{2} \frac{ P^{1, -2M+1}_{J+M-1}(\Lambda)}{P^{0, -2M}_{J+M}
(\Lambda)}
{\tilde \Lambda},
\end{equation}
with
\begin{equation}
\Lambda = 1 + 2 \bigl| \Phi_-  +  \bar \Phi_+ \bigl( 1 + |\Phi_-|^2 \bigr)
\bigr|^2
\end{equation}
and
\begin{equation}
\tilde \Lambda = 2 \bigl[ | \Phi_-|^2  ( 1 + \Phi_-  \Phi_+  +  \bar
\Phi_-   \bar \Phi_+ ) + | \Phi_+ |^2 \bigl( | \Phi_- |^4 -1\bigr)
\bigr].
\end{equation}

Thus, in the spin-${1/2}$ representation, we get
\begin{equation}
{}_{\pm} \langle Z |{\mathbb J}_3 |Z \rangle_{\pm} = \frac{1}{2}
\biggl( \frac{|\Phi_- |^2 - 1}{1 + |\Phi_- |^2}\biggr).
\end{equation}

\begin{figure} \centering
\begin{picture}(31.5,21)
\put(0,0){\framebox(31.5,21){}} \put(1,1.4){
\includegraphics[width=70mm]{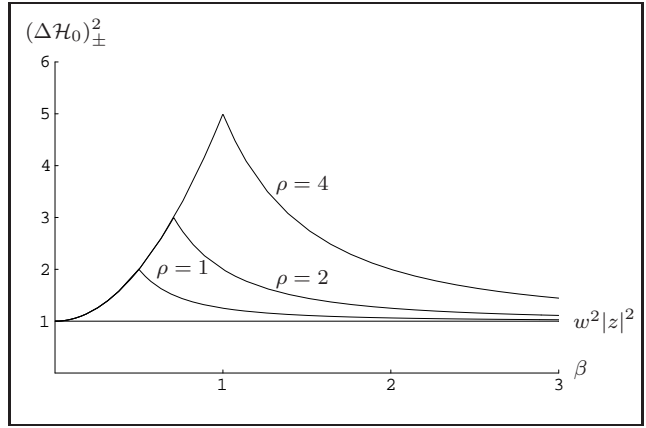}}
%\multiput(0,0)(0,5){7}{\line(1,0){45}}
%\multiput(0,0)(5,0){10}{\line(0,1){30}}
\put(27.85,2.45){\scriptsize{ $\beta$}}
\put(27.83,4.80){\scriptsize{ $w^2 {|z|}^2$}}
\put(0.5,19.25){\scriptsize{ ${( \Delta {\cal H}_0 )}^2_{\pm} $}}
\put(7.05,7.5){\scriptsize{ $\rho=1$}} \put(13,7.0){\scriptsize{
$\rho=2$}} \put(13,11.7){\scriptsize{ $\rho=4$}}
\end{picture}
\caption{Graphs of the dispersions ${ \left( {( \Delta {\cal H}_0 )
}^2 \right)}_{\pm} \equiv$(\ref{dis-ge-ne-ral}) as functions of
$\beta >0 $, $\theta=\pi$ and $\rho=1,2,4$.} \label{fig:noncano}
\end{figure}
Finally by direct computation, we find
\begin{equation}
\bigl( (\Delta {\mathcal H}_0)^2 \bigr)_\pm =  w^2|z|^2  \biggl[
1 + \rho   \frac{[16 \beta^4 \cos^2 (\theta) + \rho^2 - 8 \rho \beta^2
\cos\theta  |\cos\theta|] - R}{[16 \beta^4 \cos^2 (\theta) + \rho^2 - 8 \rho
\beta^2 \cos\theta  |\cos\theta| ] + R} \biggr], \label{dis-ge-ne-ral}
\end{equation}
where
\begin{equation}
R = \sqrt{\bigl[ 16 \beta^4 \cos^2 (\theta) - \rho^2 \cos( 2 \theta)
\bigr]^2 + \rho^4 \sin^2 (2 \theta)}.
\end{equation}
We see that, for fixed value of $\rho$,  Equation~\eqref{dis-ge-ne-ral} as a
function of
$\beta$ is symmetric around $\theta = \pi$.

\begin{figure}[h] \centering
\begin{picture}(31.5,21)
\put(0,0){\framebox(31.5,21){}} \put(1,2){
\includegraphics[width=70mm]{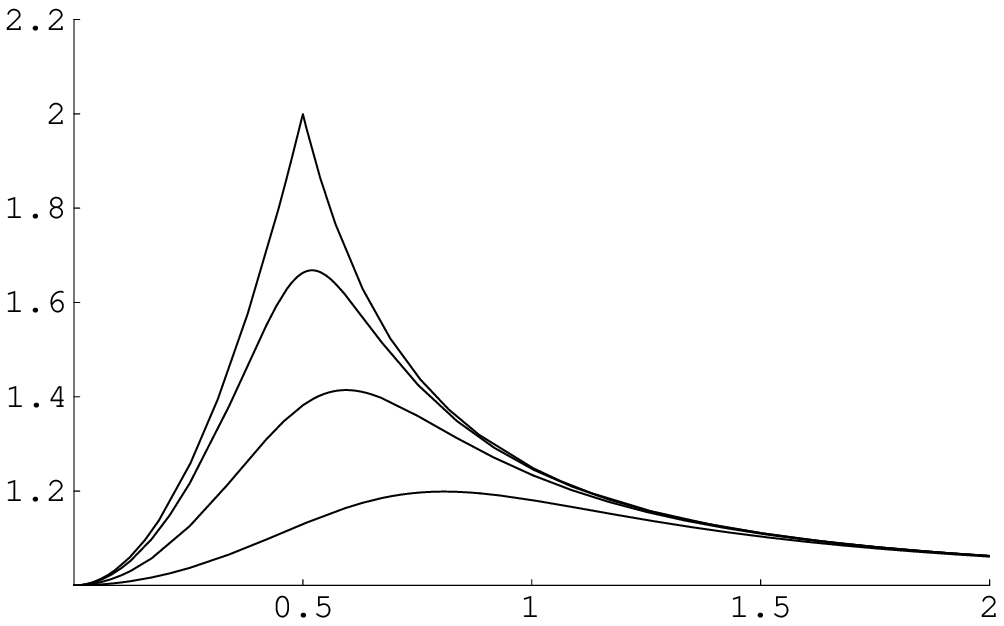}}
%\multiput(0,0)(0,5){7}{\line(1,0){45}}
%\multiput(0,0)(5,0){10}{\line(0,1){30}}
\put(15.75,0.7){\scriptsize{$\beta$}}
\put(28.35,3.1){\scriptsize{$w^2 {|z|}^2$}}
\put(8.61,3.75){\scriptsize{ $\theta={5 \pi \over 8}$}}
\put(7.71,6.0){\scriptsize{ $\theta={3 \pi \over 4}$}}
\put(7.9,8.60){\scriptsize{ $\theta={7 \pi \over 8}$}}
\put(10.94,11.9){\scriptsize{$\theta=\pi$}}
\put(1.05,19.25){\scriptsize{${( \Delta {\cal H}_0 )}^2_{\pm} $}}
\end{picture}
\caption{Graphs of the dispersions ${ \left( {( \Delta {\cal H}_0 )
}^2 \right)}_{\pm} \equiv$(\ref{dis-ge-ne-ral}) as functions of
$\beta >0 $ for  $\rho=1$, $\theta = 5 \pi / 8,  3 \pi / 4, 7 \pi
/8$  and $ \pi $.} \label{fig:phases1}
\end{figure}
Fig.~\ref{fig:noncano} shows the behaviour of the dispersions
\eqref{dis-ge-ne-ral}
as functions of $\beta > 0$ when $\theta = \pi$ and for  different values of
parameter
$\rho$. Let us notice the similarity between these curves starting from a
certain
value of $\beta$ and the curves for the canonical case showed in
Fig.~\ref{fig:dispersion1}.

Fig.~\ref{fig:phases1} shows the behaviour of the  same functions as
functions
of $\beta > 0$,  for different values of $\theta$ when $\rho=1$. We observe
that
when the angle $\theta$ is different from $\pi$ the curves have a continuous
derivative with respect to $\beta$ but,  when the angle $\theta = \pi$, the
derivative of the curve at the point $\beta = 0.5 = \sqrt\rho/2$ is not
continuous.

\section*{Acknowledgments}
 N. Alvarez M. dedicates this article to his
estimated teacher, Dr.~Luciano Laroze B., teacher at the Universidad
T\'ecnica Feder\'{\i}co Santa Mar\'{\i}a, Valpara\'{\i}so, Chile.
The research of V.~Hussin is partially supported by grants of NSERC
of Canada and FCAR du Gouvernement du Qu\'ebec.
%\newpage

\appendix\setcounter{equation}{0}\section{Algebra eigenstates associated to {\boldmath $\su(2)$}}
\label{sec-su2ae}

In this appendix we want to solve the eigenvalue equation
\begin{equation}
[ {\vec \beta} \cdot {\vec { J}} ] | \psi \rangle = [ \beta_1   J_1 +
\beta_2
J_2 + \beta _3  J_3 ]  | \psi \rangle = \Gamma | \psi \rangle,
\quad \beta_1, \beta_2, \beta_3 \in {\mathbb C},\label{epa1}
\end{equation}
where $ J_1$, $J_2 $ and  $J_3 $ are the $\su(2)$ generators which have
already
been given in Section~\ref{sec-angular}. The eigenvalue equation
\eqref{epa1}
can
also be written as
\begin{equation}
[\beta_-   J_+  + \beta_+  J_- + \beta _3  J_3]  | \psi \rangle = \Gamma |
\psi \rangle \label{epa2-t},
\end{equation}
where $J_1 $ and $J_2 $ have been expressed in terms of the usual operators
$J_{\pm} $
and
\begin{equation}
\beta_\pm = \frac{{\beta_1 \pm i \beta_2}}{2}.
\end{equation}
We see that Eq.~\eqref{angulaire2} is just a particular case of equation
\eqref{epa2-t}. The eigenvalue equation \eqref{epa2-t} has already been
solved by Brif~\cite{kn:Brif} by expanding the state $ | \psi \rangle $ in
the
standard coherent-state basis  \cite{kn:Pere},  introducing in this way
analytic
functions and asking for  solving a first order differential equation. Here,
we
consider a different method based on the operator algebra technique.

For $j$ fixed, we can show that \eqref{epa2-t} admits the eigenvalues
\begin{equation}
\Gamma^j_m = m b, \label{Gbpropres}
\end{equation}
with $ m=-j, \dots, j$    and $b = \sqrt{ \beta_1^2 + \beta_2^2 +
\beta_3^2}=
\sqrt{4 \beta_+ \beta_- + \beta_3^2}$. We then solve
\begin{equation}
[\beta_-   J_+ + \beta_+  J_- + \beta _3  J_3]  | \psi \rangle^j_m =
\Gamma^j_m | \psi \rangle^j_m \label{epa2},
\end{equation}
by using
\begin{equation}
| \psi \rangle^j_m = {(N^j_m)}^{-1/2}   T   | j , m \rangle, \label{fonor}
\end{equation}
where the $N^j_m $ are  normalization constants and $T$ is an operator that
has to be determined.  We take it as
\begin{equation}
 T =  \exp\biggl( - \frac{{ \tilde \theta}}{2} [e^{-i \tilde \phi}   J_+ -
e^{i
\tilde \phi}   J_- ] \biggr), \quad{\tilde \phi}, {\tilde \theta} \in
{\mathbb
C}.
\label{opt1}
\end{equation}
Inserting \eqref{fonor} with \eqref{opt1} into \eqref{epa2}, that  leads
to
\begin{equation}
[ \, \vec \beta  \cdot  \vec { J} \, ]   T  | j,m \rangle = m b   T  | j , m
\rangle .\label{varpro}
\end{equation}
Using the usual decomposition
\begin{equation}
 T = \exp\biggl( - e^{- i {\tilde \phi}}   \tan {\biggl(\frac{ {\tilde
\theta}}{
2}\biggr)}   J_+ \biggr) \quad
\exp\biggl(    \ln{  { \sec^2 ( \frac{{\tilde \theta}}{2 }\biggr) } }    J_3
\biggr)
\quad
\exp\biggl(  e^{ i {\tilde \phi}}   \tan{\biggl(\frac{{\tilde
\theta}}{2}\biggr)}
J_- \biggr) \label{decomposition}
\end{equation}
and the relations
\begin{equation}
e^{\eta  J_3}    J_\pm  e^{ - \eta  J_3} = e^{\pm \eta}   J_\pm, \quad
e^{\eta  J_\pm}  J_3  e^{ - \eta  J_\pm} =  J_3 \mp \eta  J_\pm, \quad
e^{\eta  J_\pm}    J_\mp  e^{ - \eta  J_\pm} =  J_\mp \pm 2 \eta  J_3 -
\eta^2  J_\pm,
  \label{relations}
\end{equation}
we can show that, for $\beta_+\ne 0$, $\beta_-\ne 0$ and $ b \ne 0$, we have
\begin{equation}
e^{i \tilde \phi} = \sqrt{ \frac{\beta_+}{\beta_-}} \label{eifi},
\end{equation}
and
 \begin{equation}
\frac{\tilde \theta}{2} = \arctan \biggl( \sqrt{\frac{b- \beta_3}{b +
\beta_3}} \biggr) \label{themedio}.
\end{equation}
Inserting the  results   \eqref{eifi} and  \eqref{themedio} in
 \eqref{decomposition},  we obtain
\begin{equation}
 T  = \exp\biggl( - \frac{{ 2 \beta_- }}{{b + \beta_3}} J_+\biggr)
\exp\biggl(  \ln \biggl(\frac{{2 b}}{{b + \beta_3}}\biggr)  J_3\biggr)
\exp\biggl( \frac{{ 2 \beta_+}}{{b +  \beta_3}}  J_-\biggr).  \label{opt2}
\end{equation}

The original form  \eqref{opt1}  of the $T$ operator allows us to look
easily  for the
special cases studied in  \cite{kn:Puri,kn:Pere} and
in the preceding sections while the form  \eqref{opt2}  allows to calculate
directly the explicit form of the eigenstates \eqref{fonor}. Indeed, the
first  relation \eqref{relations} allows us to pass the exponential term
$\exp\bigl(  \ln (2 b/(b + \beta_3))  J_3\bigr)$ to the right
in \eqref{opt2} and this without changing  essentially the operator action
on the pure states $|j,m\rangle$ because $|j,m \rangle$ is an eigenstate of
the
operator $J_3$. Thus, in equation \eqref{fonor}, we can replace the
operator
$T$ by
the operator
\begin{equation}
 T_{\mathrm{eff}} = {\biggl(\frac{b}{\beta_+}\biggr)}^{j + m}
\sqrt{\frac{{(j+m)!
(j-m)!}}{(2j)!}}
\exp\biggl( - \frac{{ 2 \beta_- }}{{b + \beta_3}} J_+\biggr)
\exp\biggl(\frac{ \beta_+}{b } J_-\biggr) \label{teff},
\end{equation}
such that
\begin{equation}
| \psi \rangle^j_m = {({\widetilde N}^j_m)}^{-1/2}   T_{\mathrm{eff}}   | j
, m
\rangle, \label{fonor-eff}
\end{equation}
where ${\widetilde N}^j_m $ are new normalization constants.
Redefining the summation indices, we get \beqa \nonumber | \psi
\rangle^j_m
        &=&  {({\widetilde N}^j_m)}^{-1/2} \sum_{u=-j}^{j}
\sqrt{\frac{{(j+u)!
(j-u)!}}{(2j)!}}   {\biggl(\frac{b}{\beta_+}\biggr)}^{j + u}  \\
        &\times& \frac{(j+m)!}{(j-u)!} \sum_{n=0}^{j+u} (-1)^n
\frac{(j-u+n)!}{{ n! (m-u+n)! (j+u-n)!}} {\biggl( \frac{{ (1 -
{\beta_3/b}) }}{2}\biggr)}^n | j , u \rangle. \eeqa We also have an
expression in terms of the Jacobi polynomials (see~\cite{kn:Vile}):
\begin{equation}
| \psi \rangle^j_m =  {({\widetilde N}^j_m)}^{-1/2} \sum_{u=-j}^{j}
\sqrt{\frac{{(j+u)!
(j-u)!}}{(2j)!}}   {\biggl(\frac{b}{\beta_+}\biggr)}^{j + u}
P^{-u+m,-u-m}_{j+u} \biggl( \frac{\beta_3}{b}\biggr)  | j , u \rangle,
\label{fonjaco}
\end{equation}
which is the result obtained by Brif~\cite{kn:Brif}.

For the special case where $\beta_+ = 0$,   $\beta_3\ne 0$  so that, in
connection with
 \eqref {Gbpropres}, we have $b = \beta_3$, we find the operator
\begin{equation}
 T_{\mathrm{eff}} = \exp\biggl(-\frac{\beta_-}{\beta_3} J_+\biggr).
\label{teff-b+=0}
\end{equation}
The eigenstates are
\begin{equation}
| \psi \rangle^j_m = (C^j_m)^{- 1/2} \sum_{u=m}^{j}
\sqrt{\frac{(j+u)!}{(j-u)!}}   \frac{1}{(u-m)!}  \biggl( -  \frac{
\beta_-}{\beta_3}\biggr)^{u-m}  | j,u \rangle,
\end{equation}
and become the standard CS of $\SU(2)$  \cite{kn:Pere} when $m =-j$.

For the special case where $\beta_- = 0$, $\beta_3\ne 0$, we have similar
results.
Indeed, the new operator $T_{\mathrm{eff}}$  is
\begin{equation}
 T_{\mathrm{eff}} = \exp\biggl(  \frac{\beta_+}{\beta_3 }  J_-\biggr)
\label{teff-b-=0}
\end{equation}
and the eigenstates write
\begin{equation}
| \psi \rangle^j_m = (C^j_m)^{-1/2}  \sum_{u=-j}^{m}
\sqrt{\frac{(j-u)!}{(j+u)!}} \frac{1}{(m-u)!}     \biggl(  \frac{
\beta_+}{\beta_3}\biggr)^{u-m}  | j,u \rangle,
\end{equation}
which become the standard CS of $\SU(2)$ \cite{kn:Pere} when $m=j$.

Now for the case $\beta_+ =0 $ and  $\beta_3 = 0$   ($\beta_- =0 $  and
$\beta_3 = 0$),
the only normalizable solution
is $|j,-j \rangle$ $\bigl(|j,j\rangle \bigr)$. For  $\beta_+ = \beta_- =0$ and
$\beta_3 \ne 0 $,  the AES are evidently the pure states $ | j , m \rangle$.

Finally, the degenerate case $b=0$ leads to the solution
$ | \psi \rangle^j_{-j} = ( C^j_{-j})^{- 1/2}  T_{\mathrm{eff}} | j
, - j \rangle$  with $ T_{\mathrm{eff}}= \exp(  {-2 (\beta_- /\beta_3)}
J_+ ) $, that is the standard CS of $\SU(2)$.

The mean value of  $ J_3$ in the states \eqref{fonjaco} has already been
calculated by Brif~\cite{kn:Brif}. We have
\begin{equation}
\langle  J_3 \rangle^j_m = \frac{{ j Y + m (S_+ -S_-)}}{{S_+ S_-}} -
\frac{{(j + |m|) Y t}}{{S_+^2 S_-^2}} \Omega,
\end{equation}
where
\begin{equation}
S_\pm = 1 + {\biggl| \frac{{2 \beta_-}}{{\beta_3 \mp b}} \biggr|}^2, \quad t
=
{\biggl|\frac{b}{\beta_+} \biggr|}^2,  \quad
 Y = S_+ S_- - S_+ - S_-
 \end{equation}
 and
\begin{equation}
\Omega = \frac{{P_{j-|m|-1}^{(-2j,1)}  (1- (2t/S_+ S_-))}}{
{P_{j-|m|}^{(-2j-1,0)} (1- (2t/S_+ S_-))}}, \quad {\rm if } \
|m|< j;
\qquad   \Omega = 0, \quad  {\rm if } \  |m|=j .
\end{equation}

\setcounter{equation}{0}\section{Resolution of a first order system
of differential equations} \label{sec-auno}

Let us recall that a realization \cite{kn:Pere} of the Fock space
${\mathcal F}_b= \bigl\{ | n\rangle, n=0,1,2,\dots \bigr\}$ of
energy  eigenstates of the harmonic oscillator as a space $\mathfrak
H$ of analytic functions $f (\zeta)$ is obtained by expanding this
function in the basis of analytic functions $\{ \varphi_n (\zeta) =
{\zeta^n/\sqrt{n !}}, n=0,1,2, \dots \}$, that is
\begin{equation}
f(\zeta) =  \sum_{n=0}^\infty c_n \varphi_n (\zeta) = \sum_{n=0}^\infty c_n
\frac{\zeta^n}{ \sqrt{n!}}, \quad \zeta\in {\mathbb C}. \label{fdezeta}
\end{equation}
The scalar product is
\begin{equation}
( f_1 , f_2 ) = \int_{C}   \bar f_1 (\zeta)  f_2 (\zeta)  e^{-
{|\zeta|}^2} \frac{{d\zeta \, d{\bar \zeta}}}{2 \pi i }, \quad
\forall    f_1 , f_2 \in {\mathfrak H}, \label{norma}
\end{equation}
the integral being extended to the complex plane. The action of the
creation $ a^\dagger$ and annihilation $ a$ operators on the
${\mathfrak H}$ space is then given by
\begin{equation}
 a^\dagger \equiv \zeta, \quad  a \equiv \frac{d}{d \zeta}.
\end{equation}
The eigenvalues  equation \eqref{evalpaa} thus becomes a first order
differential
equation
\begin{equation}
\frac{1}{\sqrt2} \biggl( (1 + \lambda) \frac{d}{d \zeta} + (1-\lambda) \zeta
\biggr) f(\zeta) = \beta  f(\zeta), \label{equ-diff-harm}
\end{equation}
for which normalized solutions are obtained for $\lambda \ne - 1$. The
general
solution
of \eqref{equ-diff-harm}  is
\begin{equation}
f(\zeta) = f(0) \exp\biggl( \frac{2 \sqrt2 \beta \zeta - (1-\lambda)
{\zeta}^2}{
2 (1 + \lambda)} \biggr).
\end{equation}
With respect to the scalar product \eqref{norma}, the normalization constant
$f(0)$ is computed by imposing
\begin{equation}
\int_C \bigl|f(\zeta)\bigr|^2 e^{- {|\zeta|}^2} \frac{{d\zeta \,  d{\bar
\zeta}}}{2 \pi i }
= 1,
\end{equation}
and we find the normalized solution of \eqref{equ-diff-harm} as
\begin{equation}
f_ (\zeta) = \bigl(1 - |\eta_1|^2 \bigr)^{1/4} \exp\biggl( - \frac{1}{2}
\biggl[ \frac{{|\eta_2|}^2 - \Remot ({\bar \eta}_1 \eta_2^2)}{1 -
{|\eta_1|}^2 }
\biggr] \biggr) \exp\biggl( \eta_2 \zeta - \frac{\eta_1}{2} {\zeta}^2
\biggr),
\end{equation}
with
\begin{equation}
\eta_1 = \frac{(1-\lambda)}{(1+\lambda)}= \delta e^{i \phi} \quad {\rm and}
\quad \eta_2= \frac{\sqrt2 \beta}{(1+\lambda)}= \frac{\beta}{\sqrt2} (1 +
\delta e^{i \phi}).
\end{equation}
This corresponds to  the states \eqref{solcompose} after normalization.

Now we are concerned  with  the algebra eigenstates satisfying the equation
 \eqref{promo} in the Fock space $\mathcal F \equiv$ \eqref{Fock-space}.
A realisation of $\mathcal F$ can be easily given from the preceding
considerations and the expression \eqref{fondaa} of a state $|\psi \rangle$
for a fixed $j$. Indeed, we have
\begin{equation}
\psi^j_m (\zeta) = \langle \zeta ; j , m |\psi \rangle
\end{equation}
and the eigenvalue equation \eqref{promo} then becomes a system of
first order differential equations \beqa  \nonumber \biggl(\alpha _-
\frac{d}{d\zeta} +   \alpha_+ \zeta + \alpha_3 \biggr) \psi^j_m
(\zeta)
        + \bigl[\beta_-  \sqrt{ (j - m + 1)( j + m)}  \psi^j_{m - 1}
(\zeta)     \\
         + \ \beta_+  \sqrt{ (j + m + 1)( j - m )}  \psi^j_{m + 1}
(\zeta)  + \beta _3 m \psi^j_m (\zeta) \bigr] = \beta \psi^j_m
(\zeta), \label{syseq} \eeqa where $j$ is fixed but $m$ takes the
values $-j, \dots, j$. Let us now solve this system by first
introducing  the differential operator
\begin{equation}
L = \alpha _- \frac{d}{d\zeta} +   \alpha_+ \zeta + \alpha_3 - \beta
\end{equation}
and, second, defining the vector
\begin{equation}
\Psi = \begin{pmatrix}\psi^j_{-j} \cr \psi^j_{-j+1} \cr \vdots \cr
\psi^j_{j-1} \cr \psi^j_{j}\cr \end{pmatrix}. \label{gen-solu}
\end{equation}
The system \eqref{syseq} thus becomes a matrix differential system
\begin{equation}
L \Psi  = - A  \Psi, \label{reduction}
\end{equation}
with $A$ a $(2j+1) \times (2j+1)$ matrix given by
\begin{equation}
A =\begin{pmatrix}
      -j \beta_3  & \sqrt{2j} \beta_+ &  0 & 0 & \dots & 0 \cr
\sqrt{2j} \beta_- & (-j+1) \beta_3 & \sqrt{(2j-1) 2} \beta_+ & 0 &
\dots & 0 \cr 0 & \sqrt{(2j-1) 2} \beta_- &  (-j+2) \beta_3 &
\sqrt{(2j-2) 3}\beta_+& \dots & 0 \cr \vdots & \vdots& \vdots &
\ddots & \dots & \vdots \cr 0 & 0 & \sqrt{(2j-2) 3} \beta_- & (j -
2) \beta_3 & \sqrt{(2j-1) 2} \beta_+ & 0 \cr 0 & 0 & 0 &
\sqrt{(2j-1) 2} \beta_- & (j -1) \beta_3 & \sqrt{2j} \beta _+ \cr 0
& 0 & 0 & 0 & \sqrt{2j} \beta_-& j \beta_3 \cr\end{pmatrix}.
\end{equation}

If we can find a nonsingular matrix $S$ that diagonalizes $A$ on the form
$D=S^{-1} A S$
where
\begin{equation}
D= \diag  ( \lambda^j_{-j},   \lambda^j_{-j+1}, \dots, \lambda^j_j ),
\end{equation}
the system  \eqref{reduction}  will reduce to
\begin{equation}
L  \tilde \Psi = - D  \tilde \Psi, \quad  \tilde \Psi = S^{-1} \Psi.
\label{sysdiag}
\end{equation}
Thus, for $\alpha_- \ne 0$, the direct integration of \eqref{sysdiag} will
lead to
\begin{equation}
\tilde \psi^j_m = \tilde \psi^j_m (0) \exp \biggl( \frac{\beta - \alpha_3 -
\lambda^j_m}{
\alpha_-} \zeta - \frac{\alpha_+}{2 \alpha_-} \zeta^2 \biggr)
\end{equation}
and the general solution $\Psi$  will be  obtained as
\begin{equation}
\begin{pmatrix}\psi^j_{-j}  \cr \psi^j_{-j+1}  \cr \vdots \cr \psi^j_{j-1} \cr
\psi^j_{j}\cr \end{pmatrix} = S
\begin{pmatrix}\tilde \psi^j_{-j} \cr \tilde \psi^j_{-j+1} \cr \vdots \cr \tilde
\psi^j_{j-1} \cr \tilde \psi^j_{j}\cr \end{pmatrix} = \sum_{m=-j}^j
\tilde \psi^j_m (0) \exp\biggl( \frac{\beta - \alpha_3 - \lambda^j_m
}{ \alpha_-} \zeta - \frac{\alpha_+}{2 \alpha_-} \zeta^2 \biggr)
\begin{pmatrix}S_{-j,  m} \cr S_{-j+1,  m} \cr \vdots \cr S_{j-1,  m} \cr S_{j,
m}\cr\end{pmatrix}, \label{casbdiff0}
\end{equation}
where $S$ is assumed to be on the form:
\begin{equation}
S= \begin{pmatrix}  S_{-j,-j} &  S_{-j,-j+1}  &  \dots & S_{-j,j-1}
& S_{-j,j} \cr S_{-j+1,-j} &  S_{-j+1,-j+1}  &  \dots & S_{-j+1,j-1}
& S_{-j+1,j} \cr \vdots & \vdots& \dots & \vdots & \vdots \cr
S_{j-1,-j} &  S_{j-1,-j+1}  &  \dots & S_{j-1,j-1} & S_{j-1,j} \cr
S_{j,-j} &  S_{j,-j+1}  &  \dots & S_{j,j-1} & S_{j,j}
\cr\end{pmatrix}.
\end{equation}
Computing the eigenvalues of $A$, we find that we have to distinguish two
cases, i.e.\
the one with $b = \sqrt{ 4 \beta_+ \beta_- + \beta_3^2} \ne 0$
and the one with $b= 0$. For the first case $b \ne 0$, all eigenvalues are
different and
given by
\begin{equation}
\lambda^j_m = m b, \quad m= -j, \dots, j.
\end{equation}
The system is diagonalizable and the general solution is given by
 \eqref{casbdiff0} with
\begin{equation}
S_{u,  m} = {\sqrt\frac{(j+u)!(j-u)!}{(2j)!}} \biggl(
\frac{b}{\beta_+}\biggr)^{j+u}
P^{-u+m, -u-m}_{j+u} \biggl( \frac{\beta_3}{ b}\biggr), \quad   u = -j,
\dots, j,
\label{solu1-A-non-de}
\end{equation}
when $\beta_- \ne 0$, $\beta_+ \ne 0$ and $\beta_3 \ne 0$,
 \begin{equation}
S_{u,  m} = {\sqrt\frac{(j-u)!}{(j+u)!}} \frac{1}{(m-u)!} \biggl(
\frac{\beta_+}{
\beta_3}\biggr)^{u-m}, \quad -j \le u \leq m, \qquad S_{u,  m} =0, \quad m <
u \le j,
 \label{solu2-A-non-de}
\end{equation}
when $\beta_-=0$, $\beta_+ \ne 0$ and $\beta_3\ne 0$ and
\begin{equation}
S_{u,  m} = {\sqrt\frac{(j+u)!}{(j-u)!}} \frac{1}{(u-m)!} {\biggl(-
\frac{\beta_-}{
\beta_3}
\biggr)}^{u-m}, \quad m \le u \leq j, \qquad S_{u,  m} =0, \quad -j \le u <
m,
 \label{solu3-A-non-de}
\end{equation}
when $\beta_- \ne 0$, $\beta_+ =0$ and $\beta_3\ne 0$.

In the Fock space representation,  the  solutions \eqref{casbdiff0} with
\eqref{solu1-A-non-de}, \eqref{solu2-A-non-de} and \eqref{solu3-A-non-de}
correspond, apart from an superfluous change of notation,  exactly to the
states \eqref{fon1} with $T_{\mathrm{eff}}$ given by \eqref{teff},
\eqref{teff-b+=0} and \eqref{teff-b-=0} respectively.

For the  second case $b=0$,   the  matrix $A$  can  not be diagonalized. We
could use the Jordan form or start from the differential equation system
again and  include this condition. Taking the second way, we can express the
$\psi^j_m (\zeta)$ components  on the form
\begin{equation}
\psi^j_m (\zeta) = \exp\biggl[ - \frac{\alpha_+}{2 \alpha_-} {\zeta}^2 +
\frac{(\beta - \alpha_3 - m \beta_3)}{\alpha_-} \zeta \biggr]  \tilde
\psi^j_m (\zeta), \label{par-suite}
\end{equation}
and insert  these in equation  \eqref{syseq}.  We get to the
following system: \beqa \alpha _- \frac{d}{d\zeta} \tilde \psi^j_m
(\zeta) + \beta_-  \sqrt{ (j - m + 1)( j + m)} e^{\beta_3
\zeta/\alpha_- } \tilde \psi^j_{m - 1}
(\zeta) \nonumber \\
 +  \ \beta_+  \sqrt{ (j + m + 1)( j - m )}  e^{-\beta_3
\zeta/\alpha_- } \tilde \psi^j_{m + 1} (\zeta) =0,
\label{syseq-tilde} \eeqa when $m=-j, \dots, j$. By handling these
equations suitably we can, for example, obtain an ordinary
differential equation of the $2j+1$  order for $\tilde \psi^j_{-j}
(\zeta)$, namely:
\begin{equation}
\biggl[\prod_{-j}^j \biggl( \frac{d}{d\zeta} - \mu^j_m\biggr) \biggr] \tilde
\psi^j_{-j} (\zeta) =0,
\end{equation}
where
\begin{equation}
\mu^j_m = - j \frac{\beta_3}{\alpha_-} + m \frac{b}{\alpha_-}.
\end{equation}
When $b=0$, we have $2j+1$ equal roots.  This means that the solutions for
$\tilde  \psi^j_{-j} (\zeta)$  take the form:
\begin{equation}
\tilde \psi^j_{-j} (\zeta) = \exp\biggl( \frac{-j \beta_3 \zeta}{
\alpha_-}\biggr)  \sum_{q=0}^{2j} A_q {\zeta}^q. \label{jmoinsj}
\end{equation}
Then,  we can insert \eqref{jmoinsj}  in  \eqref{syseq-tilde}   and thus
obtain,  in an iterative way,  all solutions $\tilde \psi^j_m (\zeta)$  and
thereafter,  using \eqref{par-suite}, all solutions $\psi^j_m (\zeta)$.

For example, in the case  $\beta_+=\beta_3=0$ and $\beta_-\ne 0$, we have
\begin{equation}
\tilde \psi^j_{-j} (\zeta) = \psi^j_{-j} (0),
\end{equation}
i.e.\  a constant and, consequently,  by integrating one by one the
equations of the system \eqref{syseq-tilde},  we obtain
\begin{equation}
\tilde \psi^j_{m} (\zeta) = \sum_{k=0}^{j+m}  {\biggl( - \frac{\beta_-}{
\alpha_-}\biggr)}^k
\frac{{\zeta}^k}{k!} \sqrt{\frac{(j+m)! (j-m+k)!}{(j-m)! (j+m-k)!}}
\psi^j_{m-k} (0),
\end{equation}
when $ m=-j, \dots, j$.  The general solution  \eqref{gen-solu}  is
then given by \beqa  \lefteqn{\Psi
        = \exp\biggl[ - \frac{\alpha_+}{2 \alpha_-} {\zeta}^2 +
\frac{(\beta - \alpha_3)}{ \alpha_-} \zeta \biggr]}\nonumber\\
 && \times          \sum_{m=-j}^{j} \psi^j_{m} (0) \left[\sum_{k=0}^{j-m}
\frac{{(-1)}^k}{k!} \sqrt{\frac{(j-m)! (j+m+k)!}{(j-m-k)!}}
{\biggl(\frac{\beta_-}{ \alpha_-}\biggr)}^k
{\zeta}^k\begin{pmatrix}0 \cr \vdots \cr 0 \cr 1 \cr 0 \cr \vdots
\cr 0 \cr \end{pmatrix} \right], \label{solu-dege}
\end{eqnarray}
where, in each sum,  the $1$  in the vector column  is placed in the
$(j+m+k+1)$ row. We thus obtain the $(2j+1)$ independent
solutions of the system of differential equations.

In the Fock space representation, we can show that the independent solutions
given by equation \eqref{solu-dege} correspond, apart from a superfluous
change of notation, to the states  \eqref{solb++b3=0}.  In the case
$\beta_-=\beta_3=0$  with $\beta_+ \ne 0$, following a similar procedure,
one finds the expression \eqref{solb--b3=0}.

Finally, when $\beta_+$, $\beta_-$, $\beta_3 \ne 0$, by inserting
\eqref{jmoinsj}   in   \eqref{syseq-tilde} and  ordering the
independent solutions with respect to  the arbitrary constants
$A_q$, one finds: \beqa  \nonumber \Psi (\zeta)
         = \exp\biggl[ - \frac{\alpha_+}{2 \alpha_-} {\zeta}^2 +
\frac{(\beta - \alpha_3)}{ \alpha_-} \zeta \biggr]\qquad \qquad
\qquad \qquad \qquad \qquad \qquad \qquad \qquad \qquad  \
\\ \times \sum_{q=0}^{2j} A_q  \left[ \sum_{k=0}^q {(-1)}^k  {q \choose
k}  \frac{(2j-k)!}{(2j)!} {\zeta}^{q-k} {\biggl( \frac{\alpha_-
}{\beta_+}\biggr)}^k\biggl[ \frac{d^k}{d{\vartheta}^k}
\sum_{r=0}^{2j} \sqrt{ \frac{(2j)!}{ (2j-r)! r!}} \vartheta^r
\biggr]
\begin{pmatrix} 0 \cr \vdots \cr 1 \cr \vdots \cr 0 \cr \end{pmatrix} \right],
\eeqa where $\vartheta = {\beta_3/2 \beta_+} = - {2
\beta_-/\beta_3}$ and in each sum,  the $1$  in the vector column is
placed in the $r+1$ row. In the Fock space representation,  these
solutions, with a slight change of notation, correspond to
Eq.~\eqref{solb+b-b3=0}.

\end{document}